\documentclass[12pt]{article}
\usepackage{amsfonts}
\usepackage{amsmath, amssymb}
\usepackage{youngtab}
\usepackage{hyperref}
\usepackage[dvips]{color}
\usepackage{psfrag}
\usepackage{feynmp}
\usepackage[pdftex]{graphicx}
\Yboxdim{5pt}

\newcommand{\ra}{\rightarrow}

\newcommand{\be}{\begin{equation}}
\newcommand{\ee}{\end{equation}}
\newcommand{\ba}{\begin{eqnarray}}
\newcommand{\ea}{\end{eqnarray}}
\newcommand{\bi}{\begin{itemize}}  
\newcommand{\ei}{\end{itemize}}

\newcommand{\Tr}{{\rm Tr}}

\newcommand{\ep}{\epsilon}

\newcommand{\aslash}[1]{\,\,{\raise.15ex\hbox{/}\mkern-12mu #1}}
\newcommand{\bslash}[1]{\,\,{\raise.15ex\hbox{/}\mkern-11mu #1}}
\newcommand{\cslash}[1]{\,\,{\raise.15ex\hbox{/}\mkern-10mu #1}}
\newcommand{\dslash}[1]{\,\,{\raise.15ex\hbox{/}\mkern-9mu #1}}

\renewcommand{\bar}{\overline}

\newcommand{\CN}{{\mathcal N}}
\newcommand{\CM}{{\mathcal M}}
\newcommand{\CS}{{\mathcal S}}

\usepackage{color}

\renewcommand{\title}[1]{\vbox{\center\LARGE{#1}}\vspace{5mm}}
\renewcommand{\author}[1]{\vbox{\center\large#1}\vspace{5mm}}
\newcommand{\address}[1]{\vbox{\center\em#1}}

\begin{document}
\bibliographystyle{utphys}
\begin{fmffile}{graphs}

\begin{titlepage}
\begin{center}
\vspace{5mm}
\hfill {\tt 
}\\
\vspace{20mm}

\title{
Line operators in supersymmetric gauge theories\\
 and the 2d-4d relation
}
\vspace{7mm}

Takuya Okuda
\vskip 6mm
\address{
University of Tokyo, Komaba\\
Meguro-ku, Tokyo 153-8902, Japan
}

\end{center}

\vspace{10mm}
\abstract{
\noindent
Four-dimensional gauge theories with $\mathcal N=2$ supersymmetry admit half-BPS line operators.
We review the exact localization methods for analyzing these operators.
We also review the roles they play in the relation between four- and two-dimensional field theories, and explain how the two-dimensional CFT can be used to obtain the quantitative results for 4d line operators.
This is a contribution to the special LMP volume on the 2d-4d relation, edited by J. Teschner.
}
\vfill

\end{titlepage}

\tableofcontents

\section{Introduction}

Gauge theory, a fundamental description of nature in our current understanding of particle physics, remains a central subject in theoretical physics.
Any quantum field theory with gauge fields possesses a set of universal observables, namely Wilson-'t Hooft line operators, also known as loop operators.
The Wilson loop $\text{Tr}\,P \exp( i\oint A)$
exhibits an area law in a confining vacuum.
A magnetic analog, the 't Hooft loop, is a disorder operator defined by a singular boundary condition of the gauge field.
A Higgs phase can be characterized by a 't Hooft loop obeying an area law.
More generally, the behavior of mixed Wilson-'t Hooft operators can be used to classify the vacuum structures of gauge theories \cite{'tHooft:1977hy}.
Quantitative understanding of these operators in a non-abelian gauge theory such as QCD is an important open problem.

Four-dimensional theories with extended supersymmetry admit BPS line operators, which represent infinitely massive BPS particles.
While they have no known role as order parameters for low-energy physics, the BPS line operators serve as useful probes of various dualities.
BPS Wilson loops in $\mathcal N=4$ super Yang-Mills theory were first introduced in \cite{Maldacena:1998im,Rey:1998ik}, and for many years they were studied mostly in the context of the AdS/CFT correspondence.
In \cite{Erickson:2000af} it was found that the perturbative ladder diagram contributions to the expectation value of a half-BPS circular Wilson loop in $\mathcal N=4$ theory with $SU(N)$ gauge group reproduce the large 't Hooft coupling result from AdS.
The ladder diagram contributions can be neatly packaged into a Gaussian matrix model, and the authors of \cite{Erickson:2000af} conjectured that the matrix model computes the Wilson loop vev in the large $N$ limit.
Based on a conformal anomaly, \cite{Drukker:2000rr} further conjectured that the agreement should hold to all orders in $1/N$ and in the 't Hooft coupling.
The agreement was finally proved to be exact in the paper \cite{Pestun:2007rz}, where general $\mathcal N=2$ theories were also treated.

This article reviews the recent developments in the study of BPS line operators in 4d $\mathcal N=2$ gauge theories.
There are two main ideas: localization and the 2d-4d correspondence.
The former, whose modern version was invented by Pestun \cite{Pestun:2007rz} building upon earlier works \cite{Witten:1988ze,Nekrasov:2002qd}, can be applied to line operators in various geometries to obtain exact results.
The latter, in particular the AGT correspondence \cite{Alday:2009aq}, can be used to compute the expectation values of line operators by 2d CFT techniques.

This article is organized as follows.
In Section \ref{sec:charges} we will review the definition of Wilson-'t Hooft line operators and the classification of charges, as well as their counterparts in two dimensions.
In Section \ref{sec:localization} we will review the localization methods applied to line operators.
Section \ref{sec:CFT-line} is devoted to explaining the 2d CFT techniques used to compute 4d observables involving line operators.
The line operators exhibit interesting algebraic structures, which are closely related to a quantization of the Hitchin moduli space.
These matters will be reviewed in Section \ref{sec:algebra}.
The appendix summarizes some relevant facts.

Our emphasis is on the intrinsic UV dynamics of 4d line operators.
The exact computation of disorder line operators, which was made possible by localization and was inspired by the 2d-4d relation, is a remarkable progress.
The 2d theories themselves also display very rich physics.
Other review articles in this volume discuss closely related subjects from different angles.

\section{Charges of line operators}
\label{sec:charges}
In this section we review the classification of BPS line operators in an $\mathcal N=2$ gauge theory with gauge group $G$.
We begin by considering all line operators allowed by the Lie algebra of $G$ and the matter content.
We will also explain the basic correspondence between line operators and closed curves on a Riemann surface.
Then we will review the recent progress on the discrete choice one must make to fully specify a quantum field theory, and how it relates to the spectrum of line operators.

\subsection{Definition and charges of 4d line operators}
\label{sec:line-def}

Let us recall some basics of Lie algebras and set notation.
(See for example the appendix of \cite{Gukov:2006jk} for a useful summary.)
We denote by $\mathfrak t$ the Cartan subalgebra of $G$, and by $\mathfrak t^*$ the dual of $\mathfrak t$.
The roots and weights of $G$ take values in $\mathfrak t^*$, and generate the root lattice $\Lambda_\text{r}$ and the weight lattice $\Lambda_\text{w}$ respectively.
We have $\Lambda_\text{r}\subset \Lambda_\text{w}$.
We define the coroot lattice $\Lambda_\text{cr}\subset \mathfrak t$ to be the dual of $\Lambda_\text{w}$, and the coweight lattice $\Lambda_\text{cw}\subset \mathfrak t$ to be the dual of $\Lambda_\text{r}$.

Wilson line operators that preserve some supersymmetry were first introduced in the context of AdS/CFT correspondence \cite{Maldacena:1998im,Rey:1998ik}.
For an $\mathcal N=2$ gauge theory on $\mathbb R^4$, let us focus on the half-BPS Wilson operators along a straight line or a circle, defined as \begin{equation}\label{eq:Wilson-def}
W_R=\text{Tr}_R P \exp\left[\oint (iA + \text{Re}\,\phi\, ds)\right]\,,
\end{equation}
where $\phi$ is a complex scalar in the vector multiplet and $ds$ is the line element determined by the metric.
There exist more general curves and scalar couplings that preserve some amount of supersymmetry; their classification may be possible by extending the methods of \cite{Dymarsky:2009si}.
Charges of supersymmetric Wilson operators are classified by irreducible representations $R$, or equivalently their highest weights $w\in \Lambda_\text{w}$.
Physically, Wilson operators represent the worldline of an electrically charged BPS particle with infinite mass.

The magnetic analogue, 't Hooft operators, were originally introduced to classify the low-energy behavior of non-conformal gauge theories \cite{'tHooft:1977hy}, and represent the trajectory of an infinitely heavy magnetic monopole in spacetime.
These operators were generalized to preserve a half of supersymmetry in \cite{Kapustin:2005py}.
A 't Hooft operator is a {\it disorder operator}, meaning that it is defined by a singular boundary condition on the fields in the path integral.
In the present case, we define the operator $T(B)$ by demanding that we integrate over the field configurations with Dirac monopole singularities
\begin{equation}\label{eq:thooft-singularity}
  F\sim\frac{B}{4}\epsilon_{ijk}\frac{x^i}{r^3} dx^k\wedge dx^j
=-\frac{B}{2}\sin\theta d\theta\wedge d\varphi\,,
\quad
\phi\sim i\frac{B}{2r}\,.
\end{equation}
In writing this, we assumed that the theta angles of the gauge theory are zero; if they are not we need to excite $\text{Re}\,\phi$ as well as electric components of the field strength on top of~(\ref{eq:thooft-singularity}), for the gauge Noether charge to vanish \cite{Witten:1979ey} and supersymmetry to be preserved.
We have introduced $(x^1,x^2,x^3)$ and $(r,\theta,\varphi)$, locally-defined Cartesian and spherical coordinates in the directions orthogonal to the trajectory.
The 't Hooft operator preserves the same set of supercharges as the Wilson operator (\ref{eq:Wilson-def}) when placed along the same curve.
The magnetic charge $B$ is constrained by the Dirac quantization condition.
Namely, the gauge potential $A\sim -(B/2)(1-\cos\theta)d\varphi$
has a Dirac string singularity along $\theta=\pi$.
For the Dirac string to be unphysical the matter fields must be single-valued.
Thus, if we denote by $\langle\ ,\,\, \rangle$ the natural pairing between coweights and weights, the magnetic charge $B$ must satisfy
\begin{equation}\label{eq:dirac-cond}
\langle B,w\rangle\in \mathbb Z 
\end{equation}
for the highest weight $w\in \Lambda_{\rm w}$ of any irreducible representation in which a matter field transforms.
The coweights $B$ satisfying (\ref{eq:dirac-cond}) form a lattice $\Lambda_{\rm m}$, the dual of the lattice generated by roots and the weights of matter representations.

More generally, dyonic operators, or mixed Wilson-'t Hooft operators, are classified by pairs $(B,w)\in \Lambda_{\rm m}\times\Lambda_{\rm w}$ modulo the action of the Weyl group.
The weight $w$ may be interpreted as the highest weight of a representation corresponding to the Wilson loop for the subgroup unbroken by $B$, for an appropriate choice of representative~$(B,w)$.

\begin{figure}[bt]
  \centering
 \includegraphics[scale=0.6]{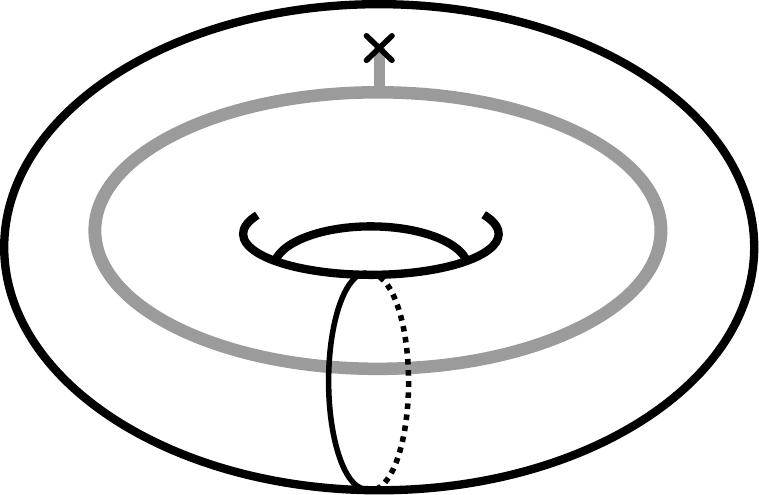}
\hspace{20mm}
 \includegraphics[scale=0.6]{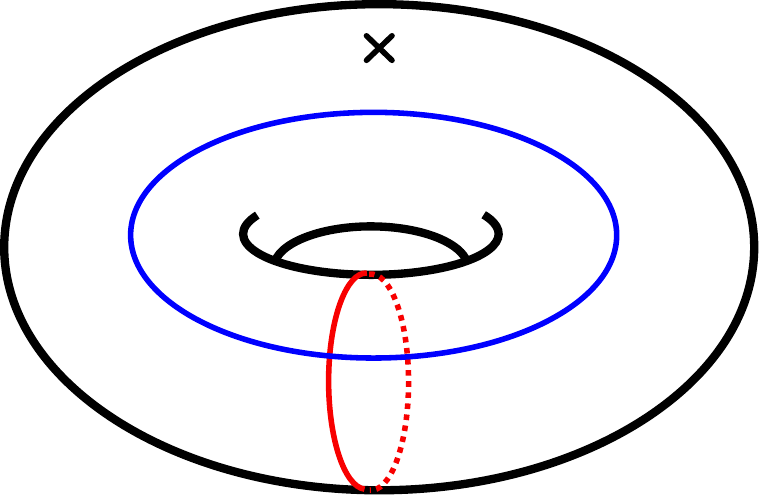}
\\
(a) \hspace{61mm} (b)
  \caption{
(a) A one-punctured torus $C_{1,1}$ is decomposed into a pair of pants with one leg degenerate.
A trivalent graph $\Gamma$ is drawn on the decomposed surface.
(b) 
The red curve, a pants leg, has $(p,q)=(0,1)$ and corresponds to the minimal Wilson operator in $\mathcal N=2^*$ theory.
The blue curve with $(p,q)=(1,0)$ corresponds to the minimal 't Hooft operator.
}
  \label{fig:torus}
\end{figure}

It is illuminating to consider the class $\mathcal S$ theories of type $A_1$,
reviewed in Appendix \ref{app:class-S} and \cite{G13}.
A weakly coupled description of such a theory is specified by a choice of pants decomposition and a trivalent graph $\Gamma$ drawn on $C_{g,n}$.
(See Figure \ref{fig:torus}(a) for an example.)
The universal covering of the gauge group $G$ is $\tilde G=SU(2)^{3g-3+n}=:\prod_{i}SU(2)_i$, where $i=1,\ldots,3g-3+n$ labels the internal edges of $\Gamma$.
The external edges labeled by $e=3g-2+n,\ldots, 3g-3+2n$ correspond to the factors in the flavor group $G_\text{F}=SU(2)^n=:\prod_e SU(2)_e$.
In addition to the electric and magnetic charges for $G$, it turns out to be convenient to allow line operators to have the magnetic charges for $G_\text{F}$ 
by coupling the theory, via the Cartan of $G_\text{F}$,
to non-dynamical gauge multiplets of the form (\ref{eq:thooft-singularity}).
Thus we consider the coweights $\vec p=(p_1,\ldots,p_{3g-3+2n})\in 
\Lambda_{\rm cw}(G\times G_\text{F})=
\mathbb Z^{3g-3+2n}$ of the extended group
and the weights $\vec q=(q_1,\ldots,q_{3g-3+n})
\in\Lambda_\text{w}(G)=\mathbb Z^{3g-3+n}$ of the gauge group.
Each trivalent vertex of the graph $\Gamma$ corresponds to a half-hypermultiplet with its scalars $\Phi_{jkl}$ transforming in the trifundamental representation of $SU(2)_j\times SU(2)_k\times SU(2)_l$, where $j,k,l$ correspond to either a gauge or flavor symmetry and do not need to be distinct.
For $\Phi_{jkl}$ to be single-valued around a Dirac string, the coweight must satisfy
\begin{equation}\label{eq:classS-Dirac}
  p_j+p_k+p_l\in 2\mathbb Z\,.
\end{equation}
The conditions (\ref{eq:classS-Dirac}), imposed for all triplets $(j,k,l)$ corresponding to vertices, define the lattice $\Lambda_{\rm m}\subset \Lambda_{\rm cw}$.
We also identify the charges related by the Weyl group $\mathbb Z_2^{3g-3+2n}$.
Thus the charges of the line operators in this theory are classified by the set of integers $(\vec p,\vec q)$, subject to the conditions (\ref{eq:classS-Dirac}) for each trivalent vertex, with the charges identified if they are related by the Weyl group action, {\it i.e.}, $(p_i,q_i)\mapsto (-p_i,-q_i)$ for some internal edge $i$ or $p_e\mapsto -p_e$ for an external edge $e$.
Identification by the Weyl group action is equivalent to requiring that $p_i,p_e\geq 0$, and also that $q_i\geq 0$ if $p_i= 0$.
Ordinary Wilson-'t Hooft operators, without non-dynamical fields involved, correspond to $(\vec p,\vec q)$ with $p_e=0$ for all external edges~$e$.

\subsection{Correspondence of charges and curves}
\label{sec:charge-curve}

We are ready to describe the basic 2d/4d relation for the charges of line operators in the $A_1$-theories \cite{Drukker:2009tz}.
We fix one pants decomposition and use it to describe all homotopy classes of closed curves on $C_{g,n}$ without self-intersection, as well as homotopy classes of arcs connecting punctures without self-intersection. 
We allow the curve $\gamma$ to have multiple components, but we assume that no component is homotopic to a point or a curve arbitrarily close to a puncture.
Let $\gamma_1$, \dots, $\gamma_{3g-3+n}$ be  pairwise disjoint connected curves without self-intersection whose complement is a pants decomposition of $C_{g,n}$; these are known as pants legs.
Also let $\gamma_{3g-3+n+1}$, \dots, $\gamma_{3g-3+2n}$ be simple closed curves near the punctures.  
In order to describe the correspondence, we define the intersection number $p_j=\#(\gamma\cap \gamma_j)$ for $1\leq j\leq 3g-3+2n$ to be the minimum of the number, without a sign, of intersection points as $\gamma$ and $\gamma_j$ vary among non-self-intersecting curves in their respective isotopy classes.
We also need a notion of twisting number $q_j$ for $1\leq j \leq 3g-3+n$.
Roughly, $q_j$ counts how many times $\gamma$ winds around in a direction parallel to $\gamma_j$.
We refer the reader to \cite{Drukker:2009tz} for a precise definition.
The crucial mathematical fact for the correspondence is the following theorem.

\vspace{5mm}

\noindent {\bf Dehn's Theorem.}
Let $C_{g,n}$ be an oriented punctured Riemann surface of genus $g$ and negative Euler characteristic with $n$ punctures.
Let us define a map 
\begin{equation}
\label{map-Dehn}
\gamma \mapsto (\#(\gamma\cap \gamma_j); q_j) \in (\mathbb Z_{{}\ge0})^{3g-3+2n}\times \mathbb Z^{3g-3+n}
\end{equation}
which assigns, to each isotopy class of closed curves without self-intersection or arcs connecting punctures without self-intersection, its intersection number $p_j=\#(\gamma\cap \gamma_j)$ with $\gamma_j$ ($1\le j\le 3g-3+2n$) and its twisting number $q_j$ with respect to $\gamma_j$ ($1\le j\le 3g-3+n$).
Note that the intersection and twisting numbers depend only on the homotopy class of $\gamma$.
With this definition the map is injective, and its image in (\ref{map-Dehn}) is
\begin{equation}
  \begin{aligned}
\{ (p_1,p_2,\ldots,&\, p_{3g-3+2n};q_1,q_2,\ldots, q_{3g-3+n})  
\nonumber
\\
& |\ \text{ if } p_j=0 \text{ then } q_j\ge0,
\text{ and } p_{j}+p_{k}+p_{l} \in 2\mathbb Z
\label{eq:Dehn-classification}
\\ &\text{ when }  \gamma_{j}  \cup  \gamma_{k}  \cup  \gamma_{l}
\text{ is the boundary of a pair of pants} \}.
\nonumber
  \end{aligned}
\end{equation}
The integers $p_j$, $q_j$ are known as the  Dehn-Thurston parameters of $\gamma$.
\vspace{5mm}

It is immediate to recognize (\ref{eq:Dehn-classification}) as the same data that classify the line operator charges in the $A_1$-theory corresponding to $C_{g,n}$.
This is the most basic 2d-4d correspondence involving line operators \cite{Drukker:2009tz}.
An example in Figure \ref{fig:torus}(b) shows curves on the one-punctured torus corresponding to line operators in the $SU(2)$ $\mathcal N=2^*$ theory.
The rules for the action of the modular groupoid on the Dehn-Thurston parameters are explicitly known \cite{MR743669}.
Since this action is interpreted as S-duality according to the 2d/4d correspondence, the line operators with charges $(\vec p,\vec q)$ are believed to transform according to the same rules.

\subsection{Spectrum of line operators and discrete theta angles}
\label{sec:spectrum}

The specification of a gauge theory, before picking a spacetime geometry, requires several discrete choices.
The choice can be phrased in terms of line operators \cite{Gaiotto:2010be,Aharony:2013hda}.

On $\mathbb R^4$ for some purposes one considers all ``line operators'' that are allowed by the Lie algebra of the gauge group and the matter content.
This is useful in the classification of massive phases of a gauge theory \cite{'tHooft:1977hy} by representations of the 't Hooft commutation relation
\begin{equation}\label{thooft-com}
W\cdot T=e^{2\pi i/N} T\cdot W
\end{equation}
for fundamental Wilson ($W$) and 't Hooft ($T$) loops.
Here  the gauge group has the Lie algebra of $SU(N)$, and we consider two closed curves $C_W$ and $C_T$ that are contained and Hopf-linked in a constant time slice.
We place $T$ on $C_T$ while we displace $W$ infinitesimally from $C_W$ forward and backward in time, so that the two sides of (\ref{thooft-com}) arise as operator products that are differently time-ordered.
Massive phases such as confining and Higgs vacua arise as representations of  (\ref{thooft-com}).
Even though $C_W$ and $C_T$ are linked within the three-dimensional slice, $C_T$ and displaced $C_W$ are, for dimensional reasons, not linked in the ambient spacetime.
The relation $W\cdot T\neq T\cdot W$ means that the two operators cannot both be genuine line operators.
If the gauge group is $SU(N)$, $W$ is a genuine loop operator that is invariant under all gauge transformations, even when it is placed along a homotopically non-trivial curve.
On the left hand side of (\ref{thooft-com}) we can link $W$ with the surface swept by the Dirac strings that extend from $T$ in the future time direction.
Then $W\cdot T$ picks up the phase $e^{2\pi i/N}$, relative to $T\cdot W$, from the holonomy around the Dirac string.
One cannot continuously deform one configuration to the other without $W$ hitting the Dirac sheet of $T$.
Thus $T$ is a boundary of a surface operator defined by the Dirac sheet.
For gauge group $SU(N)/\mathbb Z_N$ with a zero theta angle, $T$ is a genuine line operator and $W$ is a boundary of a surface operator \cite{Kapustin:2014gua}.

For a general gauge group $G$ and two pairs of charges $(B_1,w_1),(B_2,w_2)\in \Lambda_{\rm cw}\times \Lambda_{\rm w}$,  the product of two would-be loop operators acquires a phase
\begin{equation} \label{phase-genera}
\exp\left(2\pi i  \langle B_2, w_1\rangle- 2\pi i \langle B_1, w_2\rangle\right)
\end{equation}
when one operator moves around along a surface that links the other.
The correlation function of two genuine loop operators is well-defined only if the phase vanishes, in which case they are {\it mutually local}.
In a consistent theory line operators must be mutually local, and their spectrum must be {\it maximal} in the sense that one cannot add more line operators without violating mutual locality \cite{Gaiotto:2010be,Aharony:2013hda}.
The center $Z$ of the universal covering $\tilde G$ of $ G$, and its dual $Z^*$, are isomorphic to $\Lambda_{\rm cw}/\Lambda_{\rm cr}$ and $\Lambda_{\rm w}/\Lambda_{\rm r}$, respectively.
The phase (\ref{phase-genera}) depends only on the elements $z_1,z_2\in Z\times Z^*$ that correspond to the two loop operators.
A maximal mutually local spectrum then translates to a maximal isotropic subgroup of $Z\times Z^*$.

The authors of \cite{Aharony:2013hda} showed that the choice of a maximal coisotropic subgroup is equivalent to the choice of what they call {\it discrete theta angles}.
These parameters give, in the Euclidian path integral, non-trivial phases dependent on the topological classes of gauge bundles.
Correspondence with line operators arises because the discrete magnetic flux through the $\mathbb S^2$ surrounding a line operator induces, combined with discrete theta angles, a discrete electric charge, much as in the usual Witten effect \cite{Witten:1979ey}.
For $A_1$ theories of class $\mathcal S$ corresponding to a Riemann surface $C$ with no puncture, the discrete choice corresponds to a maximal coisotropic subgroup $\Delta$ of $H^1(C,\mathcal C)$, where $\mathcal C$ is the center of the simply connected group with Lie algebra $\mathfrak g$ \cite{Gaiotto:2010be,Tachikawa:2013hya}.

\section{Exact results for line operators by localization}
\label{sec:localization}

Line operators with electric and magnetic charges constructed in the previous section fit nicely the framework of supersymmetric localization \cite{Witten:1988ze}.
In this section we review the exact localization computation of loop/line operators in 4d ${\mathcal N=2}$ theories on $\mathbb S^4$ and other geometries.
Since details on the $\mathbb S^4$ localization can be found in \cite{Pestun:2007rz} as well as in the review article \cite{PV}, we restrict ourselves to the bare basics and focus on the features specific to 't Hooft operators.
The results will be successfully matched with 2d computations in Section~\ref{sec:comparison}.

\subsection{Localization for Wilson loops on $\mathbb S^4$ }
\label{sec:localization-wilson}

The general procedure in a localization computation consists of the following steps \cite{Pestun:2007rz}.
\begin{enumerate}
\item 
Pick a supercharge $\mathcal Q$ that annihilates the operator one wants to evaluate. 
 If $\mathcal Q^2$ contains terms that vanish on shell, {\it i.e.}, vanish by the equation of motion, add auxiliary fields so that such terms do not appear.
Then $\mathcal Q^2$ is a linear combination of bosonic symmetry generators.

\item Choose a $\mathcal Q^2$-invariant functional $V$ such that the bosonic terms of $\mathcal Q\cdot V$ are positive-semidefinite.
Add $t \mathcal Q\cdot V$ to the action where $t$ is a constant.
Find the saddle points of the path integral $\int e^{-S_\text{cl}-t \mathcal Q\cdot V}$ in the limit $t\rightarrow +\infty$, in other words find the configurations such that $\mathcal Q\cdot V=0$.

\item 
Evaluate the classical action $S_\text{cl}$ and the inserted operator at the saddle point.

\item \label{item:evaluate}
Compute the fluctuation determinant at the saddle points.  This involves gauge-fixing and the inclusion of ghost fields.  Either expand fields in the eigenmodes of kinetic operators, or use the equivariant index theorem to compute the determinant.

\item Sum and integrate the above contributions over all the saddle points.
\end{enumerate}

It is possible to define $\mathcal N=2$ non-conformal supersymmetry on $\mathbb S^4$, and the steps above were carried out in \cite{Pestun:2007rz} to compute the partition functions and the Wilson loop vevs for $\mathcal N=2$ gauge theories on $\mathbb S^4$.

 One of the key steps is the computation of the fluctuation determinant.
This is the ratio $\det \Delta_o/(\det\Delta_e)^{1/2}$, where $(\Delta_e,\Delta_o)$ are the differential operators acting on bosons and fermions in $\mathcal Q\cdot V$.
 On $\mathbb S^4$ we choose $V=\sum (\text{fermion})\cdot \overline{\mathcal Q(\text{fermion})}$, and  $(\Delta_e,\Delta_o)$ can be expressed in terms of simpler differential operators in $V$.
 Supersymmetry implies many cancellations among the eigenvalues, and one can show that the determinant is given by
\begin{equation}\label{eq:det-reduced}
\frac{\det \Delta_o}{(\det\Delta_e)^{1/2}}=
\left(
 \frac{\text{det}_{\text{coker}\mathcal D}\mathcal Q^2}{ \text{det}_{\text{ker}\mathcal D} \mathcal Q^2}
\right)^{1/2}\,,
\end{equation}
where $\mathcal D$ is a differential operator in $V$, and we recall that $\mathcal Q^2$ is a sum of bosonic symmetry generators.
Schematically,
\begin{equation}\label{eq:Q-squared}
  \mathcal Q^2\sim J+R+a+m\,,
\end{equation}
where $J$, $R$, $a$, and $m$ generate an isometry, an R-symmetry rotation, a gauge transformation, and a flavor symmetry transformation.
Despite the huge cancellations the fluctuation determinant (\ref{eq:det-reduced}) is still an infinite product and takes the form $\prod_j w_j^{c_j/2}$ with $c_j=\pm 1$.
The weights $w_j$ and the signs $c_j$ can be read off from the equivariant index
\begin{equation}
   \text{ind}\, \mathcal D\equiv
\Tr_{\text{ker} \mathcal D} e^{ \mathcal Q^2}
-
\Tr_{\text{coker} \mathcal D} e^{ \mathcal Q^2}=\sum c_j e^{w_j}\,,
\end{equation}
which can be computed by the Atiyah-Singer index theory.
In particular, the fixed point formula expresses the index as a sum of contributions from the fixed points of the isometry $J$.
Thus the fluctuation determinant for each saddle point configuration, computed by the index theory, naturally factorizes into the contributions from the fixed points, namely the north and south poles of $\mathbb S^4$.

At the north pole, the differential operator $\mathcal D$ acts on the vector multiplet fields as the linearization of the anti-self-duality equations, which govern instantons on $\mathbb C^2$:
\begin{equation}\label{eq:SD}
\Omega^1(\text{ad}\,E)
\stackrel{
\text{\raisebox{-1mm}[0mm][0mm]{$D^\dagger \oplus (1+*) D$}}
}{
\xrightarrow{\hspace*{19mm}}} 
\Omega^0
(\text{ad}\,E)
\oplus
\Omega^{2+}(\text{ad}\,E)\,.
\end{equation}
On the hypermultiplet $\mathcal D$ acts as the Dirac operator.
The structure at the south pole is similar, with anti-instantons replacing instantons.

Non-perturbative contributions arise as small instantons and anti-instantons localized at the north and south poles respectively.
More precisely, these are the $\mathcal Q^2$-fixed points on the instanton moduli space.
Let us denote by ${Z}_{\text{1-loop}}^\text{pole}\left(ia, im_f\right)$ the north pole contribution to the fluctuation determinant in the topologically trivial backgrounds.
The variable $a$, taking values in the Cartan subalgebra $\mathfrak{t}$, parametrizes the saddle point configurations and is identified with the background value of a vector multiplet scalar $\text{Re}\,\phi$.
The parameters $m_f$ denote the masses of matter hypermultiplets.
A topologically non-trivial configuration at the north pole contributes the universal factor ${Z}_{\text{1-loop}}^{\text{pole}}\left(ia, im_f\right)$, accompanied by an extra rational function of $a$ and $m_f$.
The sum of the rational functions over the invariant instanton configurations is the instanton partition function $Z_\text{inst}$ \cite{Nekrasov:2002qd} with the omega deformation parameters specialized to the values $\epsilon_1=\epsilon_2=1/r$, where $r$ is the radius of $\mathbb S^4$.
The contributions from the south pole have a similar structure.

Factorization of the determinants implies that the total partition function takes the form
\begin{equation}\label{eq:S4-vev}
Z_{\mathbb S^4}=\langle 1\rangle_{\mathbb S^4}= \int_{\mathfrak t} da \, \left|Z^\text{pole}(a)\right|^2 \,,
\end{equation}
where the integral is taken over the Cartan subalgebra $\mathfrak t$, and
\begin{equation}\label{eq:Z-pole}
Z^{\text{pole}}(a)=  Z_{\text{cl}}
\left(ia,\tau \right){Z}^\text{pole}_\text{1-loop}\left(ia, im_f\right)Z_{\text{inst}}\left(ia, r^{-1}+ i m_f;r^{-1},r^{-1}; \tau\right)\,.
\end{equation}
We factorized the classical part by hand:
$e^{-S_\text{cl}}=|Z_\text{cl}|^2$.
The precise expressions of various factors in (\ref{eq:Z-pole}) in a similar convention can be found in \cite{Gomis:2011pf}.
The instanton partition function $Z_\text{inst}(a,m_f;\epsilon_1,\epsilon_2;\tau)$ defined in \cite{Nekrasov:2002qd} arises as a sum of the rational functions over $\mathcal Q$-invariant instanton configurations localized at each pole.
To compute the vev of the Wilson loop defined by (\ref{eq:Wilson-def}) with an integral along the equator, we only need to evaluate it in each saddle point as indicated in step \ref{item:evaluate} above:
\begin{equation}\label{eq:Wilson-S4}
\langle W_R\rangle_{\mathbb S^4}=
\int_{\mathfrak t} da \, \left|Z^\text{pole}(a)\right|^2 
\text{Tr}_R e^{2\pi i r a}\,.
\end{equation}
In particular, this reduces to the Gaussian matrix model for $\mathcal N=4$ theory, proving the conjecture \cite{Erickson:2000af,Drukker:2000rr} mentioned in the introduction.

\subsection{Instanton/monopole correspondence}
\label{sec:inst-mono}

We now review a similar localization calculation for a 't Hooft operator \cite{Gomis:2011pf}.
A nice technical tool is a correspondence between singular monopoles on $\mathbb R^3$
and $U(1)$-invariant instantons on
a Taub-NUT space, discovered by Kronheimer \cite{Kronheimer:MTh}.
For our purposes, it is enough to specialize to the single-center Taub-NUT space with metric
\begin{equation}
 ds^2=V (
d\rho^2+
\rho^2d\theta^2+\rho^2 \sin^2\theta d\varphi^2
)+V^{-1}(d\psi+\omega)^2\,,
\label{TN-metric}
\end{equation}
where $V=l+1/(2\rho)$, $\omega=(1/2) (1-\cos\theta)d\varphi$, $l>0$ is a constant, and $(\rho,\theta, \varphi)$ are the polar coordinates for $\mathbb R^3$.
This is a circle fibration over the flat $\mathbb R^3$.
The variable $\psi$ has periodicity $2\pi$ in our convention.
From the three-dimensional fields $(A,\Phi)$ with singularities
\begin{equation}
  A\sim- \frac {B}2 (1- \cos\theta) d\varphi\,,
\quad
\Phi \sim \frac{B}{2\rho}
\end{equation}
near the origin,
we construct a four-dimensional gauge connection
\begin{equation}
  \mathcal A \equiv  
g
\left(A+\Phi \frac{d\psi+\omega}{V}\right)
g^{-1}
-i g d g^{-1}
\label{calA}
\end{equation}
and its curvature $\mathcal F=d\mathcal A+i\mathcal A\wedge\mathcal A$.
Here $g= e^{i B \psi}$ is a singular gauge transformation.
The singularities in $A$ and $\Phi$ cancel out in (\ref{calA}), and we obtain a smooth four-dimensional gauge field $\mathcal A$.

The four-dimensional field $\mathcal A$ is invariant 
under the $U(1)_K$ action $\psi \ra \psi  +{\rm const}.$,
which acts on the circle fiber as well as the gauge bundle.
The correspondence states that the Bogomolny equations
\begin{equation}\label{eq:Bogo}
D_i\Phi=\frac12 \epsilon_{ijk} F_{jk}\qquad (i,j,k=1,2,3)
\end{equation}
on $\mathbb R^3$ are equivalent to the anti-self-dual equations
\begin{equation}
\mathcal F+  *_4\mathcal F=0\,.
\end{equation}
Since the Taub-NUT space  is isomorphic to $\mathbb C^2$ as a complex manifold, we can use instantons on $\mathbb C^2$ to perform calculations for 't Hooft operators.

\subsection{Localization for   't Hooft loops on $\mathbb S^4$}
\label{sec:localization-thooft}

\begin{figure}[tb]
\begin{center}
 \includegraphics[scale=0.4]{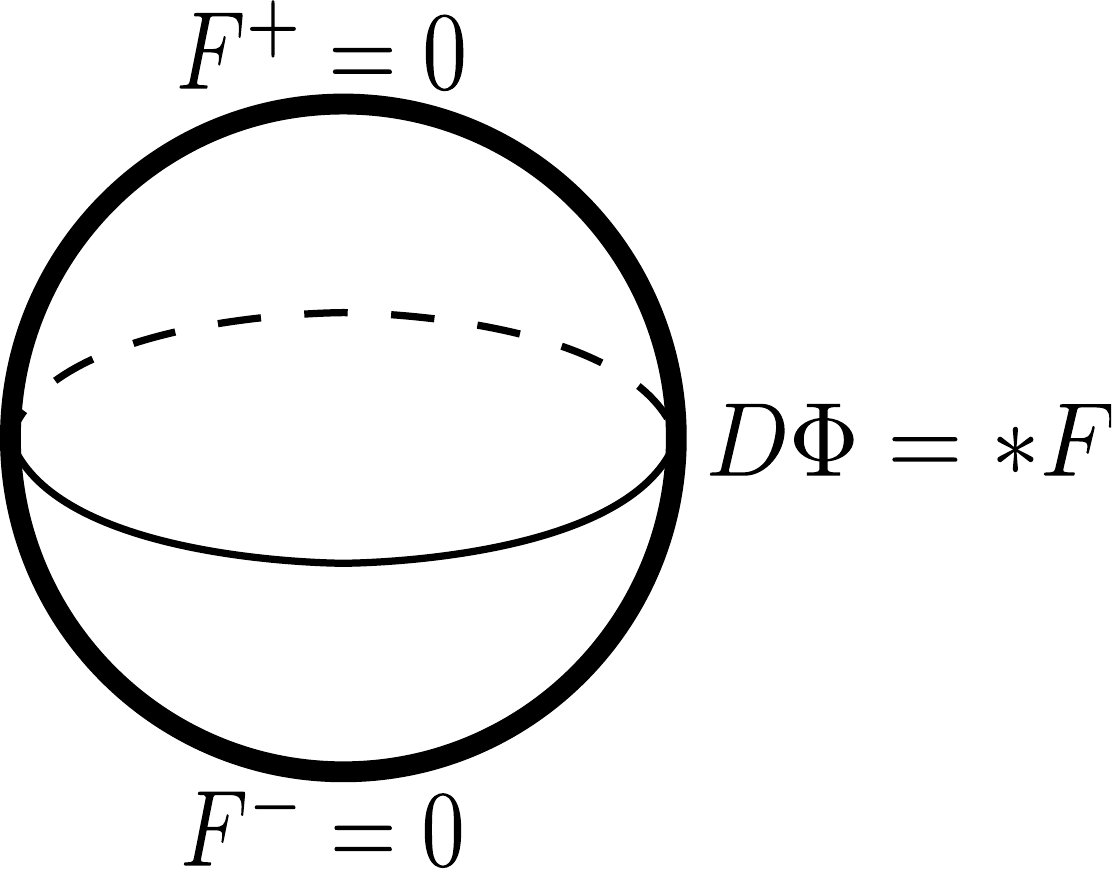}
\vspace{-5mm}
 \end{center}
 \caption{Instanton, monopole and anti-instanton field configurations.}
\label{fig:S4}
\end{figure}

Let us consider a supersymmetric 't Hooft loop $T(B)$, specified by the coweight $B$ and placed along a large circle of $\mathbb S^4$, which we refer to as the equator.
See Figure \ref{fig:S4}.
Since the 't Hooft operator is a disorder operator, we need to evaluate the path integral with the boundary conditions (\ref{eq:thooft-singularity}) which affect the saddle point configurations.
We introduce a convenient set of coordinates, in which the standard round metric on $\mathbb S^4$ of radius $r$ is given by
\begin{equation}
ds^2=
r^2 {\left(1-{|\vec x|^2\over 4r^2}\right)^2\over \left(1+{|\vec x|^2\over 4r^2}\right)^2}d\tau^2
+
{\sum_{i=1}^3dx_i^2\over \left(1+{|\vec x|^2\over 4r^2}\right)^2} \,.
\end{equation}
This is Weyl-equivalent to the metric on $\mathbb S^1\times \mathbb H^3$, where $\mathbb H^3$ is the three-dimensional hyperbolic space.
In terms of the Weyl-rescaled fields on $\mathbb S^1\times \mathbb H^3$, the only $\mathcal Q$-invariant configurations, smooth away from the operator and subject to the boundary condition (\ref{eq:thooft-singularity}), are given by
\begin{equation}\label{eq:thooft-background}
  F_{jk} = -\frac B 2 \ep_{ijk} \frac{x_i}{|\vec x|^3}\,, \quad
\text{Re}\,\phi = \frac{a}{1 + {|\vec x|^2\over 4r^2}}\,,
\quad
\text{Im}\,\phi = \frac{B}{2 |\vec x|}\,.
\end{equation}
We again assumed that $\vartheta=0$.
Note that $a\in \mathfrak t$ is the only unfixed parameter which we must integrate over.
In this background, the usual classical action diverges due to the infinite mass of the Dirac monopole.
Suitable boundary terms \cite{Giombi:2009ek,Gomis:2011pf} cancel the divergence and make the action finite.
The values of $\phi$ at the north and south poles are now shifted.
The basic effect of the 't Hooft loop is to shift the argument $a$ of $Z^\text{pole}(a)$ by $iB/2r$.

The one-loop determinant receives extra contributions.
The differential operator $\mathcal D$ in $V$ is now modified, and $\text{ind}\,\mathcal D$ receives contributions not only from the north and south poles, but also from the equator.
In the neighborhood of the equator, which can be approximated by $\mathbb S^1\times\mathbb R^3$, $\mathcal D$ acts on the vector multiplet as the differential in the complex defined on $\mathbb R^3$:
\begin{equation}
D_\text{Bogo}:
\Omega^1(\text{ad}\,E)\oplus \Omega^0(\text{ad}\,E)
\ra
 \Omega^0(\text{ad}\,E)
\oplus
\Omega^1(\text{ad}\,E)\,.
\label{eq:complex-Bogo}
\end{equation}
The arrow involves the dual of the gauge transformation and the linearization of the Bogomolny equation (\ref{eq:Bogo}).
The instanton/monopole correspondence above can be extended to the correspondence between the $U(1)_K$-invariant sections of the self-dual complex (\ref{eq:SD}) and the sections of the Bogomolny complex (\ref{eq:complex-Bogo}).
Thus the index of the latter can be obtained from that of the former by averaging over the $U(1)_K$ action.
On the hypermultiplet, $\mathcal D$ acts as $D_\text{DH}$, the Dirac operator with a coupling to the Higgs field $\Phi$.
The one-loop contribution from the equator $Z^\text{eq}_\text{1-loop}(ia,im_f,B)$ can then be read off from $\text{ind}(D_\text{Bogo})+\text{ind}(D_\text{DH}) $ by taking into account also the Fourier modes along the $\mathbb S^1$.

There are also extra non-perturbative contributions.
Recall that zero-size instantons and anti-instantons localized at the north and south poles provide non-perturbative saddle points, even though the configurations are singular.
Similarly, infinitesimal dynamical monopoles, which get attached to the Dirac monopole defining the 't Hooft loop and screen the magnetic charge, also provide non-perturbative saddle points.
The saddle point configurations are invariant under $\mathcal Q$, and hence also under $\mathcal Q^2$.
Thus the saddle points are the fixed points, with respect to a certain group action, in the moduli space $\mathcal M(B)$ of solutions of the Bogomolny equations on $\mathbb R^3$ with a Dirac singularity.

The moduli space $\mathcal M(B)$ has components labeled by  $v\in \Lambda_\text{cr}+B$ with $|v|\leq |B|$.
All the fixed points in $\mathcal M(B,v)$ take the form of the 't Hooft background (\ref{eq:thooft-background})
except that $B$ is replaced by $v$.
The classical contribution depends only on $v$
and is universal among the fixed points in $\mathcal M(B,v)$.
We also need to include the fluctuation determinant from each fixed point.
By factoring out $Z^\text{eq}_\text{1-loop}(ia,im_f,v)$,
we denote the sum of such determinants by
\begin{equation}
  Z^\text{eq}(a;B,v)\equiv
Z^\text{eq}_\text{1-loop}(ia,im_f,v)  Z^\text{eq}_\text{mono}(ia,im_f,B,v)
\equiv 
\mathop{\sum_\text{fixed points}}_{\text{in }\mathcal M(B,v)} \prod_j w_j^{c_j/2}\,.
\end{equation}
This equation defines $Z^\text{eq}_\text{mono}$.

Collecting all the contributions (see Figure \ref{fig:S4}), the result for the 't Hooft loop expectation value on $\mathbb S^4$ is
\begin{equation}\label{eq:thooft-S4}
\langle T(B)\rangle_{\mathbb S^4}=\int_{\mathfrak t} da \,\sum_v |Z^\text{pole}(a+iv/2r)|^2 Z^\text{eq}(a;B,v)\,.
\end{equation}
For example, the vev of the minimal 't Hooft loop $T\equiv  L_{1,0}$ in the $SU(2)$ $\mathcal N=2^*$ theory is given by
\begin{equation}\label{eq:2star-thooft}
\langle T\rangle_{\mathbb S^4}
=\sum_{v=\pm 1/2}
\int_{-\infty}^\infty |Z^\text{pole}(a+iv/2r)|^2
\frac{\cosh^{1/2}(\pi r(2a+m))\cosh^{1/2}(\pi r(2a-m))}{\cosh(2\pi r a)}\,.
\end{equation}
This has no bubbling contribution.
See \cite{Gomis:2011pf} for examples with non-trivial bubbling, as well as examples with dyonic charges.

\subsection{Other geometries}

\subsubsection{$\mathbb S^1\times_b\mathbb R^3$}
\label{sec:S1R3}

As we saw above, the geometry in the neighborhood of the equatorial $\mathbb S^1$ in the four-sphere is essentially $\mathbb S^1\times\mathbb R^3$.
This suggests that the contributions intrinsic to the loop operators are most naturally formulated on $\mathbb S^1\times\mathbb R^3$ itself rather than on $\mathbb S^4$.
The main advantage of  $\mathbb S^1\times\mathbb R^3$ is that we can cleanly introduce an analog of omega deformation parameter \cite{Nekrasov:2002qd}.
In particular the fundamental definition of $Z_\text{mono}^\text{eq}$, the bubbling contributions, is given in this geometry, just as the instanton partition functions are defined on flat $\mathbb C^2$ in the omega background.
While there are other geometries that admit an omega deformation, the vevs of 't Hooft operators in such geometries are expressed in terms of the quantities defined on $\mathbb S^1\times\mathbb R^3$.
Another motivation to consider the omega-deformed product $\mathbb S^1\times\mathbb R^3$ comes from the study of the IR dynamics in the same set-up \cite{Gaiotto:2010be,Ne}.

We wish to evaluate the expectation values of line operators wrapping the $\mathbb S^1$ in $\mathbb S^1\times\mathbb R^3$.
To preserve SUSY, the metric need not be a direct product; $\mathbb R^3$ may be fibered over $\mathbb S^1$.
If we regard the circle as the time direction, the line operator, an infinitely heavy particle, modifies the theory and defines a Hilbert space $\mathcal H_L(\mathbb R^3)$.
The fibration of $\mathbb R^3$ is accounted for by the insertion of the operator $e^{2\pi i b^2 J_3}$, where $J_3$ generates rotations about the 3-axis.
It should be accompanied by $e^{2\pi i b^2 I_3}$ for supersymmetry, where $I_3$ is a generator of the $SU(2)$ R-symmetry group.
We will denote such a space by $\mathbb S^1\times_b\mathbb R^3$ to indicate the twist.
Recall that an omega deformation is a SUSY-preserving modification of a theory by (equivariant) parameters $(\epsilon_1,\epsilon_2)$ for the action of the $U(1)\times U(1)$ isometry \cite{Nekrasov:2002qd}.
In our case the action $U(1)\times U(1)$ acts as the rotation of the $\mathbb S^1$ factor and a spatial rotation in $\mathbb R^3$.
The rotation commutes with supersymmetry if the ratio of the rotation angles is $\epsilon_1/\epsilon_2=b^2$.
 We can also introduce flavor symmetry generators $F_f$  and their dual variables ${\boldsymbol m}_f$, which play the role of masses.
The expectation value of a line operator (more generally the correlation function of such operators)  can be represented as a supersymmetric trace
\begin{equation}\label{eq:S1R3-trace}
\langle L\rangle_{\mathbb S^1\times_b\mathbb R^3}= \Tr_{\mathcal H_L(\mathbb R^3)} 
(-1)^{F}e^{-2\pi R H}e^{2\pi i b^2 (J_3+I_3)}
e^{-2\pi i \boldsymbol m_f F_f}\,.
\end{equation}
This quantity, in particular the one-loop and bubbling contributions $Z_\text{1-loop}^{\mathbb S^1\times_b\mathbb R^3}$ and $Z_\text{mono}^{\mathbb S^1\times_b\mathbb R^3}$,
 can be computed by localization in the same way as for the equator contributions to $\langle L\rangle_{\mathbb S^4}$.
The main difference is that in the current case the isometry can act on $\mathbb S^1$ and $\mathbb R^3$ with a variable ratio $b^2\in \mathbb R$ of rotation angles.
The result of localization for the 't Hooft operator $T_B$ is \cite{Ito:2011ea}
\begin{equation}
  \langle T_B\rangle_{\mathbb S^1\times_b\mathbb R^3}= 
\sum_v e^{2\pi i v\cdot \boldsymbol b}
Z_\text{1-loop}^{\mathbb S^1\times_b\mathbb R^3}(\boldsymbol a,\boldsymbol m_f,b;v)
Z_\text{mono}(\boldsymbol a,\boldsymbol m_f,b;B,v)\,,
\label{eq:thooft-S1R3}
\end{equation}
where $\boldsymbol a$ and $\boldsymbol b$ are respectively the vevs of $\text{Re}\,\phi$ and $\text{Im}\,\phi$ suitably rescaled and complexified by the gauge field.
Since the bubbling contribution on this geometry is most fundamental, we simply write $Z_\text{mono}\equiv Z_\text{mono}^{\mathbb S^1\times_b\mathbb R^3}$.
Indeed $Z_\text{mono}(B,v)$ serve as building blocks for the line operator correlation functions in other geometries.
In particular, it is related to the equator contribution on $\mathbb S^4$ as
$
 Z^\text{eq}_\text{mono}( a,m_f;B,v)=
 Z_\text{mono}(r a, r m_f+1/2,\lambda=1;B,v)$.
The shift in mass appears because the curved metric affects the periodicity of spinors.

For the minimal 't Hooft operator in the $SU(2)$ $\mathcal N=2^*$ theory,
\begin{equation}\label{eq:thooft-S1R3-twostar}
  \langle T\rangle_{\mathbb S^1\times_b\mathbb R^3}=
\left( e^{2\pi i \boldsymbol b}+e^{-2\pi i \boldsymbol b}\right)
\left(
\frac{
\sin\left(2\pi \boldsymbol a+\pi\boldsymbol m \right)
\sin\left(2\pi \boldsymbol a-\pi\boldsymbol m\right)
}{
\sin\left(2\pi\boldsymbol  a+\frac{\pi}2 b^2\right)
\sin\left(2\pi\boldsymbol a-\frac{\pi}2 b^2\right)
}
\right)^{1/2}\,,
\end{equation}
with $\boldsymbol a,\boldsymbol b\in\mathbb C$.
See \cite{Ito:2011ea} for more examples.

\subsubsection{$\mathbb S^1\times_b\mathbb S^3$}
\label{sec:S1S3}

Given an $\mathcal N=2$ superconformal theory, one can perform radial quantization on $\mathbb R^4$ by regarding the radial direction as time.
One can compactify this direction, and the resulting path integral, the partition function of a supersymmetric theory on $\mathbb S^1\times \mathbb S^3$, is known as the superconformal index \cite{Romelsberger:2005eg,Kinney:2005ej,RR}.
One can refine this by including line operators \cite{Dimofte:2011py}.
Line operators, wrapping the $\mathbb S^1$ and inserted at arbitrary points along a great circle of $\mathbb S^3$, preserve some common supersymmetry~\cite{Dimofte:2011py}.
In particular, if one starts with a line operator passing through the origin, by radial quantization one ends up with a line operator $L$ at the north pole of $\mathbb S^3$, and its conjugate $\bar L$ at the south pole.
Let us denote by $\mathcal H_{L,\bar L}(\mathbb S^3)$ the Hilbert space on $\mathbb S^3$ with such insertions.
The index with the line operator insertions can be represented as a supersymmetric trace
\begin{equation}
  \langle L \cdot\bar L\rangle_{\mathbb S^1\times_b\mathbb S^3}=
\text{Tr}_{\mathcal H_{L,\bar L}(\mathbb S^3)}\,(-1)^F e^{2\pi i b^2(J_L+J_R+I_3)}\prod_f \eta_f^{F_f}\,.
\end{equation}
Here $J_L$ and $J_R$ are the Cartan isometry generators for $\mathbb S^3$, $b$ plays the role of a omega deformation parameter, and $\eta_f$ are the flavor chemical potentials.
Recall also from (\ref{eq:S1R3-trace}) that $I_3$ and $F_f$ are the R- and flavor symmetry generators respectively.

This was computed in \cite{Gang:2012yr} by a hybrid method, where the one-loop contributions are computed by counting BPS states on $\mathbb S^3$, and the bubbling contributions, which we expect to be localized to the line operators and given by $Z_\text{mono}$ above, are put by hand.
The classical contribution vanishes as it should for an index.
The results agree with a prescription proposed in \cite{Dimofte:2011py}, as well as the predictions of S-duality in $\mathcal N=4$ theory.
In addition, the Wilson line operator index for $SU(N)$ $\mathcal N=4$ theory in the large $N$ limit was found to agree with the counting of fluctuation modes on the fundamental string (for the fundamental representation) and on the D5-brane (for the anti-fundamental representation).

\subsubsection{$\mathbb S^4_b$}
\label{sec:S4b}

Since the CFT side has a variable $b$ parametrizing the central charge, it was immediately recognized after the discovery of the AGT correspondence that the localization computations on $\mathbb S^4$ should be generalized.
This was done in \cite{Hama:2012bg} and is reviewed in \cite{PV}.
The metric of the new geometry, the ellipsoid $\mathbb S^4_b$, is given by $ds^2=\sum_{I=0}^4 dX_I^2$, where
\begin{equation}\label{eq:S4b}
  X_0^2+b^{-2}(X_1^2+X_2^2)+b^2(X_3^2+X_4^2)=r^2\,.
\end{equation}
This geometry has an obvious isometry $U(1)\times U(1)$, whose action commutes with supersymmetry if the ratio of rotation angles is $\epsilon_1/\epsilon_2=b^2$.
Setting $b$ to $1$ gives back the round $\mathbb S^4$ of radius $r$.
The reference \cite{Hama:2012bg} formulated $\mathcal N=2$ gauge theory (with reduced SUSY) on this geometry by introducing a background gauge field for R-symmetry  and other auxiliary fields in the $\mathcal N = 2$ supergravity multiplet.
The partition function now takes the form 
\begin{equation}\label{eq:S4b-pf}
Z_{\mathbb S^4_b}=\int da |Z^\text{pole}(a, m_f;b;\tau)|^2  
\end{equation}
 where $Z^\text{pole}(a,m_f;b;\tau)$ includes the one-loop and instanton contributions, with omega deformation parameters $(\epsilon_1,\epsilon_2)=(b,b^{-1})$.
For any fixed value of $X_0\in (-r,r)$, there are two circles along which we can place BPS loop operators:
the circle $\mathbb S^1_{(b)}$ ($X_3=X_4=0$) on the 12-plane, and another circle $\mathbb S^1_{(1/b)}$ on the 34-plane.
The computation of the Wilson loop vev and S-duality suggest that the 't Hooft loop vev is independent of $X_0$.
Let us focus on $\mathbb S^1_{(b)}$ at $X_0=0$.
The vev of the Wilson loop $W_R$ was computed in \cite{Hama:2012bg}; we just insert ${\rm Tr}_R\, e^{2\pi i b r a}$ into (\ref{eq:S4b-pf}).
We emphasize that 't Hooft loops on $\mathbb S^4_b$ have not been treated yet at the time of writing.
We expect that the essential part of computation is determined by the symmetry generated by $\mathcal Q^2$; in particular it generates the $U(1)\times U(1)$ isometry with equivariant parameters $(b,b^{-1})$.
Generalizing (\ref{eq:thooft-S4}), we should get
\begin{equation}\label{eq:S3b-thooft}
\hspace{-2mm}
  \langle T_B\rangle_{\mathbb S^4_b} \stackrel{\text{expected}}{=}
\hspace{-1mm}
 \int_{ \mathfrak t} da\,
\sum_{v}
Z^\text{eq}\left(a,m_f;b; B,v\right) \left|Z^\text{pole}\left(a+ib\frac{v}{2r},m_f;b;\tau\right)\right|^2\,,
\end{equation}
where $Z^\text{eq}\left(a,m_f;b;B,v\right):=Z^{\mathbb S^1\times_b\mathbb R^3}_\text{1-loop+mono}(i  b  r a, i b  r m_f +1/2;b;B,v)$ \cite{Ito:2011ea}.
We will compare this with the CFT computations in the next section.

\subsection{1/8 BPS Wilson loops in $\mathcal N=4$ theory and the 2d Yang-Mills}

On $\mathbb R^4$ or $\mathbb S^4$, $\mathcal N=4$ super Yang-Mills has a variety of loop operators that preserve at least one supercharge \cite{Dymarsky:2009si}.
The half-BPS Wilson loop given in (\ref{eq:Wilson-def}), where we regard $\mathcal N=4$ theory as an $\mathcal N=2$ theory with a massless adjoint hypermultiplet, is one of them.
There are different classes of line operators to which the localization method has been applied \cite{Pestun:2009nn}.

These include 1/8-BPS Wilson loops along arbitrary contours on a two-sphere.
One can also place a half-BPS 't Hooft loop that links the $\mathbb S^2$.
Certain local operators can be further inserted on $\mathbb S^2$.
Localization can be performed using the common supercharge preserved by the operators.
The results are rather different from the case in the previous subsection; the path integral has been shown to reduce, up to the assumption that the one-loop determinant is trivial, to another quantum field theory, namely the bosonic two-dimensional Yang-Mills on the $\mathbb S^2$.
The correlation functions of the operators turn out to be captured by the analogous observables in the two-dimensional theory, as conjectured in \cite{Drukker:2007yx}.
For certain combinations of the operators, the theory further reduces to multi-matrix models.
The results have been tested in a variety of ways using AdS/CFT and S-duality.
For more details, see \cite{Giombi:2009ds,Giombi:2009ek,Giombi:2012ep} and the references therein.

\section{CFT techniques for line operators }
\label{sec:CFT-line}

 In this section we review two-dimensional methods for computing the expectation values of loop operators on $\mathbb S^4_b$.
We mostly consider the  $A_1$-theories of class $\mathcal S$ and the corresponding 2d theory, namely Liouville theory.
Generalization to higher rank gauge theories and the $SU(N)$ Toda theories will be explained in Section \ref{sec:Toda}.

\subsection{Verlinde operators}
\label{sec:Verlinde}

As we saw in Section \ref{sec:charge-curve}, line operator charges in an $A_1$-theory are in a one-to-one correspondence with closed curves on the Riemann surface $C_{g,n}$.
In a conformal field theory, one can associate to any closed curve $\gamma$ an operation on conformal blocks, defined in terms of the monodromy of a degenerate field along $\gamma$.
This operation, called the {\it Verlinde operator},
 was introduced by E. Verlinde and applied to the characters (torus conformal blocks) of rational CFT's to argue that the modular S-matrix diagonalizes the coefficients in the fusion rule \cite{Verlinde:1988sn}.
Moore and Seiberg proved this conjecture, known as the Verlinde formula, by expressing the Verlinde operators in terms of the fusion and braiding moves, which are the basic ingredients for a general modular transformation of conformal blocks \cite{Moore:1988uz,Moore:1988ss}.
It turns out that the same construction works even for non-rational CFT's such as Liouville and Toda theories.
See \cite{T} for a more rigorous discussion of the Verlinde operator.
As we will elaborate in Section \ref{sec:Toda}, there is an alternative definition of a Verlinde operator as a topological defect, whose definition may be conceptually cleaner.

A degenerate field is a primary for which the Kac determinant vanishes.
As such it has a descendant that is orthogonal to all states and is decoupled \cite{Belavin:1984vu}.
In the standard parametrization of the Liouville central charge $c=1+6Q^2$ ($Q=b+b^{-1}$), primaries $V_\alpha$ have the conformal weight  $\Delta(\alpha)=\alpha(Q-\alpha)$.
The most basic degenerate fields are $V_\alpha$ with momenta $\alpha=-b/2$ and $\alpha=-1/2b$.
In view of the quantum symmetry $b\leftrightarrow b^{-1}$ of Liouville theory, it suffices to consider $V_{-b/2}$.
The condition for decoupling can be stated as 
\begin{equation}\label{eq:decoupling}
\left(  \partial^2_z + b^2 T(z)\right) V_{-b/2}(z)=0
\quad
 \text{ or equivalently} 
\quad
(L_{-1}^2+b^2 L_{-2})\cdot V_{-b/2}=0
\,,
\end{equation}
where $T(z)$ is the energy-momentum tensor and $L_n$ are the standard Virasoro generators.
Using this this, one can show that $V_{-b/2}$ and $V_\alpha$ obey the fusion rule \cite{Belavin:1984vu}
\begin{equation}
  [V_{-b/2}] [V_\alpha]=[V_{\alpha-b/2}]+[V_{\alpha+b/2}]\,.
\end{equation}
In particular, the OPE of two degenerate fields $V_{-b/2}$ contains the vacuum state $V_0=1$.

Since (\ref{eq:decoupling}) involves only a holomorphic coordinate $z$, the full correlation function with $V_{-b/2}$, as well as the corresponding conformal blocks, obeys the resulting differential equation.
In particular the conformal blocks transform linearly ({\it i.e.}, there is monodromy) when $z$ is transported along a closed curve.

We are now ready to define the Verlinde operator.
Let us consider a conformal block $\mathcal F$ specified by a trivalent graph $\Gamma$ on $C_{g,n}$.
Pick also a closed curve $\gamma$ on $C_{g,n}$, and assume that it has only one connected component.
We can define an extended conformal block $\hat{\mathcal F}(z,z_0)$ by inserting $V_{-b/2}(z)$ and $V_{-b/2}(z_0)$ at two nearby points $z$ and $z_0$, taking their OPE, and projecting onto the identity state.
Let $\Delta(\alpha)=\alpha(Q-\alpha)$ denote the conformal weight of $V_{\alpha}$.
We can recover the original block $\mathcal F$ from $\hat{\mathcal F}$ by taking the limit $z\rightarrow z_0$:
\begin{equation}
  \hat{\mathcal F}(z,z_0)
\sim
\frac{1}{(z-z_0)^{2\Delta(-b/2)}} \mathcal F \,.
\end{equation}
A priori $\hat{\mathcal F}(z,z_0)$ is defined only for $z$ close enough to $z_0$.
We can analytically continue $\hat{\mathcal F}$ and transport $z$ along a closed curve homotopic to $\gamma$, and {\it then} take the limit $z\rightarrow z_0$
\begin{equation}
  \hat{\mathcal F}(z,z_0)
\sim
\frac{c_0}{(z-z_0)^{2\Delta(-b/2)}}
\mathcal L_\gamma \cdot \mathcal F
\,.
\end{equation}
We included a universal normalization constant $c_0$, which we will specify.
The map $\mathcal F\mapsto \mathcal L_\gamma \cdot \mathcal F$ is the Verlinde operator.

As an illustration, let us take as $\hat{\mathcal F}$ a four-point conformal block with two degenerate fields at $0$ and $z$, and two primaries with momentum $\alpha$ at $1$ and $\infty$.
The block can be expressed in terms of the Gauss hypergeometric function ${}_2 F_1$:
 \begin{equation}\label{eq:block-hyper}
\hat{\mathcal F}(z,0)=
\hspace{3mm}
\raisebox{0mm}[10mm][12mm]{\parbox{30mm}{
 \begin{fmfgraph*}(30,20)
\fmfbottom{bl,br}
\fmftop{tl,tr}
\fmf{plain, width=1}{bl,b1}
\fmf{dashes}{b1,b2,b3}
\fmf{photon, width=1}{b3,br}
\fmf{phantom}{tl,t1,t2,t3,tr}
\fmffreeze
\fmf{plain,width=1}{t1,b1}
\fmf{photon,width=1}{t3,b3}
\fmfv{label=$\alpha$,label.angle=-90,label.dist=8}{bl}
\fmfv{label=$\alpha$,label.angle=180}{t1}
\fmfv{label=$-b/2$,label.angle=0}{t3}
\fmfv{label=$0$,label.angle=-90,label.dist=8}{b2}
\fmfv{label=$-b/2$,label.angle=-90,label.dist=8}{br}
 \end{fmfgraph*}
}}
\qquad
=
\frac{ (1-z)^{b \alpha }}{z^{2\Delta(-b/2)}}\,{} _2F_1\left(1+b^2,2 b \alpha ;2+2 b^2;z\right)\,.
\end{equation}
The solid lines carry a generic momentum, the wiggly ones degenerate states, and the dashed line represents the identity state.
The monodromy around $z=1$ can be computed by \href{http://functions.wolfram.com/HypergeometricFunctions/Hypergeometric2F1/17/02/07/0006/}{a hypergeometric identity} and yields a linear combination of (\ref{eq:block-hyper}) and the block with internal momentum $-b$.
One can check that the coefficient of the former is
\begin{equation}\label{eq:factor-Wilson}
\frac{e^{2\pi  ba}+e^{-2\pi  ba}}{e^{\pi i b Q}+e^{-\pi i b Q}}\,,
\end{equation}
where we defined $a$ by $\alpha=Q/2+ia$.
We now choose $c_0=1/(e^{\pi i b Q}+e^{-\pi i b Q})$.
Then the corresponding Verlinde operator is the multiplication by $e^{2\pi  ba}+e^{-2\pi  ba}$, which is the trace of a matrix in the fundamental representation of $SU(2)$.

More generally the Verlinde operator $\mathcal L_\gamma$ as defined above can be computed very explicitly by fusion and braiding moves.
These moves relate the conformal blocks assigned to two trivalent graphs that differ by a local modification.
A sequence of such modifications allows us to analytically continue the block as a function of $z$ along $\gamma$.
Since our aim is to transport a degenerate field $V_{-b/2}(z)$, we only need to implement these moves when the modification involves at least one external edge with $V_{-b/2}$.
In this special case, the fusion move is expressed in terms of a $2\times 2$ matrix $F_{s_1 s_2}$ ($s_1,s_2=\pm$).
See (\ref{eq:fexp}).
Since the fusion move, by definition, locally modifies a trivalent graph by replacing an s-channel with a t-channel, it is enough to describe its action on a four-point block:
 \begin{equation}\label{eq:fusion}
\raisebox{0mm}[10mm][12mm]{\parbox{30mm}{
 \begin{fmfgraph*}(30,20)
\fmfbottom{bl,br}
\fmftop{tl,tr}
\fmf{plain,width=1}{bl,b1,b2,b3,br}
\fmf{phantom}{tl,t1,t2,t3,tr}
\fmffreeze
\fmf{plain,width=1}{t1,b1}
\fmf{photon,width=1}{t3,b3}
\fmfv{label=$\alpha_4$,label.angle=-90,label.dist=8}{bl}
\fmfv{label=$\alpha_3$,label.angle=180}{t1}
\fmfv{label=$-b/2$,label.angle=0}{t3}
\fmfv{label=$\alpha_1-s_1\frac{b}{2}$,label.angle=-90,label.dist=8}{b2}
\fmfv{label=$\alpha_1$,label.angle=-90,label.dist=8}{br}
 \end{fmfgraph*}
}}
\quad\
=
\quad
\sum_{s_2=\pm}
F_{s_1 s_2}
\big[\begin{smallmatrix} \alpha_3 & -b/2\\ \alpha_4 & \alpha_1\end{smallmatrix}\big]
\times
\qquad
\raisebox{0mm}[10mm][12mm]{\parbox{24mm}{
\begin{fmfgraph*}(20,20)
\fmfbottom{bl,br}
\fmftop{tl,tr}
\fmf{plain,width=1}{bl,b1,b2,b3,br}
\fmf{phantom}{tl,t1,tt1,t2,tt3,t3,tr}
\fmffreeze
\fmf{plain,width=1}{t1,c2}
\fmf{photon,width=1}{c2,t3}
\fmf{plain,width=1,tension=2}{c2,b2}
\fmfv{label=$\alpha_4$,label.angle=-90,label.dist=8}{bl}
\fmfv{label=$\alpha_1$,label.angle=-90,label.dist=8}{br}
\fmfv{label=$\alpha_3-s_2\frac{b}{2}$,label.angle=-30}{c2}
\fmfv{label=$\alpha_3$,label.angle=180}{t1}
\fmfv{label=$-b/2$,label.angle=0}{t3}
\end{fmfgraph*}
}}
\end{equation}
We also use the braiding move
\begin{equation}\label{eq:braid}
\raisebox{0mm}[10mm][8mm]{\parbox{24mm}{\begin{center}
\begin{fmfgraph*}(15,15)
\fmfbottom{bl,br}
\fmftop{tl,tr}
\fmf{phantom}{bl,b1,b2,b3,br}
\fmf{phantom}{tl,t1,tt1,t2,tt3,t3,tr}
\fmffreeze
\fmf{plain,width=1}{t1,c2,t3}
\fmf{plain,width=1,tension=2}{c2,b2}
\fmfv{label=$\alpha_1$,label.angle=30}{b2}
\fmfv{label=$\alpha_2$,label.angle=180}{t1}
\fmfv{label=$\alpha_3$,label.angle=0}{t3}
\end{fmfgraph*}\end{center}}}
=e^{\pi i(\Delta(\alpha_1)-\Delta(\alpha_2)-\Delta(\alpha_3))}\times
\raisebox{0mm}[10mm][8mm]{\parbox{24mm}{\begin{center}
\begin{fmfgraph*}(15,15)
\fmfbottom{bl,br}
\fmftop{tl,tr}
\fmf{phantom}{bl,b1,b2,b3,br}
\fmf{phantom}{tl,t1,tt1,t2,tt3,t3,tr}
\fmffreeze
\fmf{plain,width=1}{t1,c2,t3}
\fmf{plain,width=1,tension=2}{c2,b2}
\fmfv{label=$\alpha_1$,label.angle=30}{b2}
\fmfv{label=$\alpha_3$,label.angle=180}{t1}
\fmfv{label=$\alpha_2$,label.angle=0}{t3}
\end{fmfgraph*}\end{center}}}
\end{equation}
where $\Delta(\alpha)=\alpha(Q-\alpha)$ is the conformal dimension of the operator $V_\alpha$ and the two exchanged vertex operators are rotated by 180 degrees.
If the rotation is in the other direction the phase is the opposite.

The curve for a spin $1/2$ Wilson operator for a gauge $SU(2)$ factor corresponds to an internal edge of the trivalent graph \cite{Alday:2009fs,Drukker:2009id}.
Indeed the action of the Verlinde operator $\mathcal L_W$ along this curve is calculated by the following sequence of moves.
\begin{eqnarray}\label{eq:Wilson-move}
&
\raisebox{0mm}[10mm][9mm]{\parbox{14mm}{\begin{center}
\begin{fmfgraph*}(10,15)
\fmfbottom{bl,br}
\fmftop{tl,tr}
\fmf{plain,width=2}{bl,b1,br}
\fmf{phantom}{tl,t1,tr}
\fmffreeze
\fmf{boson,width=.5}{tl,c1,tr}
\fmf{dashes,width=.5}{c1,b1}
\fmfv{label=$\alpha$,label.angle=-90,label.dist=8}{bl}
\fmfv{label=$\alpha$,label.angle=-90,label.dist=8}{br}
\end{fmfgraph*}\end{center}}}
\to
\raisebox{0mm}[10mm][9mm]{\parbox{24mm}{\begin{center}
\begin{fmfgraph*}(20,15)
\fmfbottom{bl,br}
\fmftop{tl,tr}
\fmf{plain,width=2}{bl,b1,b2,b3,br}
\fmf{phantom}{tl,t1,t2,t3,tr}
\fmffreeze
\fmf{boson,width=.5}{t1,b1}
\fmf{boson,width=.5}{t3,b3}
\fmfv{label=$\alpha$,label.angle=-90,label.dist=8}{bl}
\fmfv{label=$\alpha$,label.angle=-90,label.dist=8}{br}
\fmfv{label=$\alpha'$,label.angle=-90,label.dist=5}{b2}
\fmfv{label=$-\frac{b}{2}$,label.angle=180}{t1}
\fmfv{label=$-\frac{b}{2}$,label.angle=0}{t3}
\end{fmfgraph*}\end{center}}}
\to
\raisebox{-6mm}[10mm][12mm]{\parbox{19mm}{\begin{center}
\begin{fmfgraph*}(15,30)
\fmfbottom{bl,br}
\fmftop{tl,tr}
\fmf{phantom}{bl,cl,tl}
\fmf{phantom}{br,cr,tr}
\fmffreeze
\fmf{plain,width=2}{cl,c1,c3,cr}
\fmf{phantom}{tl,t1,t3,tr}
\fmf{phantom}{bl,b1,b3,br}
\fmffreeze
\fmf{boson,width=.5}{t1,c1}
\fmf{boson,width=.5}{b3,c3}
\fmfv{label=$\alpha$,label.angle=-90,label.dist=8}{cl}
\fmfv{label=$\alpha$,label.angle=-90,label.dist=8}{cr}
\end{fmfgraph*}\end{center}}}
\to
\raisebox{0mm}[10mm][9mm]{\parbox{24mm}{\begin{center}
\begin{fmfgraph*}(20,15)
\fmfbottom{bl,br}
\fmftop{tl,tr}
\fmf{plain,width=2}{bl,b1,b2,b3,br}
\fmf{phantom}{tl,t1,t2,t3,tr}
\fmffreeze
\fmf{boson,width=.5}{t1,b1}
\fmf{boson,width=.5}{t3,b3}
\fmfv{label=$\alpha$,label.angle=-90,label.dist=8}{bl}
\fmfv{label=$\alpha$,label.angle=-90,label.dist=8}{br}
\fmfv{label=$\alpha'$,label.angle=-90,label.dist=5}{b2}
\end{fmfgraph*}\end{center}}}
\to
\raisebox{0mm}[10mm][9mm]{\parbox{14mm}{\begin{center}
\begin{fmfgraph*}(10,15)
\fmfbottom{bl,br}
\fmftop{tl,tr}
\fmf{plain,width=2}{bl,b1,br}
\fmf{phantom}{tl,t1,tr}
\fmffreeze
\fmf{boson,width=.5}{tl,c1,tr}
\fmf{dashes,width=.5}{c1,b1}
\fmfv{label=$\alpha$,label.angle=-90,label.dist=8}{bl}
\fmfv{label=$\alpha$,label.angle=-90,label.dist=8}{br}
\end{fmfgraph*}\end{center}}}
\\
&
\raisebox{2mm}
\nonumber
\end{eqnarray}
The result is
\begin{equation}\label{eq:Wilson-mult}
\mathcal L_W\cdot \mathcal F= 
(e^{2\pi  ba}+e^{-2\pi  ba})
\, \mathcal F\,.
\end{equation}
Of course this agrees with (\ref{eq:factor-Wilson}) divided by $c_0$, and is the trace of an $SU(2)$ matrix.

More generally, a closed curve that traverses at least one pair of pants corresponds to a line operator with non-zero magnetic charge, and the associated Verlinde operator involves a non-trivial shift in Liouville momenta.
For example, the Verlinde operator corresponding to the minimal 't Hooft loop in $\mathcal N=2^*$ theory is given by the following moves.
$$
\hspace{6mm}
\raisebox{0mm}[10mm][12mm]{\parbox{40mm}{
\begin{fmfgraph*}(30,30)
\fmfsurroundn{o}{8}
\fmf{phantom,tension=.5}{o1,v1}
\fmf{phantom,tension=.5}{o2,v2}
\fmf{phantom,tension=.5}{o3,v3}
\fmf{phantom,tension=.5}{o4,v4}
\fmf{phantom,tension=.5}{o5,v5}
\fmf{phantom,tension=.5}{o6,v6}
\fmf{phantom,tension=.5}{o7,v7}
\fmf{phantom,tension=.5}{o8,v8}
\fmfcyclen{plain,width=2,right=0.20}{v}{8} 
\fmffreeze
\fmf{phantom}{v5,c}
\fmf{plain,width=2}{c,v1}
\fmf{phantom}{o4,o45,o5,o55,o6}
\fmffreeze
\fmf{boson,width=.5}{o45,oo,o55}
\fmf{dashes,width=.5}{oo,v5}
\fmfv{label=$\alpha$,label.angle=-90,label.dist=6}{v7}
\fmfv{label=$\alpha_e$,label.angle=-90,label.dist=6}{c}
\end{fmfgraph*}}}
\hspace{-10mm}
\to
\raisebox{0mm}[10mm][12mm]{\parbox{40mm}{
\begin{fmfgraph*}(30,30)
\fmfsurroundn{o}{8}
\fmf{phantom,tension=.5}{o1,v1}
\fmf{phantom,tension=.5}{o2,v2}
\fmf{phantom,tension=.5}{o3,v3}
\fmf{phantom,tension=.5}{o4,v4}
\fmf{phantom,tension=.5}{o5,v5}
\fmf{phantom,tension=.5}{o6,v6}
\fmf{phantom,tension=.5}{o7,v7}
\fmf{phantom,tension=.5}{o8,v8}
\fmfcyclen{plain,width=2,right=0.20}{v}{8} 
\fmffreeze
\fmf{phantom}{v5,c}
\fmf{plain,width=2}{c,v1}
\fmf{phantom}{o4,o45,o5,o55,o6}
\fmffreeze
\fmf{boson,width=.5}{o45,v4}
\fmf{boson,width=.5}{o55,v6}
\fmfv{label=$\alpha$,label.angle=-90,label.dist=6}{v7}
\fmfv{label=$\alpha'$,label.angle=180,label.dist=6}{v5}
\fmfv{label=$\alpha_e$,label.angle=-90,label.dist=6}{c}
\end{fmfgraph*}}}
\hspace{-10mm}
\to
\raisebox{0mm}[10mm][12mm]{\parbox{40mm}{
\begin{fmfgraph*}(30,30)
\fmfsurroundn{o}{8}
\fmf{phantom,tension=.5}{o1,v1}
\fmf{phantom,tension=.5}{o2,v2}
\fmf{phantom,tension=.5}{o3,v3}
\fmf{phantom,tension=.5}{o4,v4}
\fmf{phantom,tension=.5}{o5,v5}
\fmf{phantom,tension=.5}{o6,v6}
\fmf{phantom,tension=.5}{o7,v7}
\fmf{phantom,tension=.5}{o8,v8}
\fmfcyclen{plain,width=2,right=0.20}{v}{8} 
\fmffreeze
\fmf{phantom}{v5,c}
\fmf{plain,width=2}{c,v1}
\fmf{phantom}{o4,o45,o5,o55,o6}
\fmf{phantom}{o8,o85,o1}
\fmffreeze
\fmf{boson,width=.5}{o55,v6}
\fmf{boson,width=.5}{o85,v8}
\fmfv{label=$\alpha$,label.angle=-90,label.dist=6}{v7}
\fmfv{label=$\alpha'$,label.angle=180,label.dist=6}{v5}
\fmfv{label=$\alpha''$,label.angle=-60,label.dist=4}{v1}
\fmfv{label=$\alpha_e$,label.angle=-90,label.dist=6}{c}
\end{fmfgraph*}}}
\hspace{-6mm}
\to
\raisebox{0mm}[10mm][12mm]{\parbox{40mm}{
\begin{fmfgraph*}(30,30)
\fmfsurroundn{o}{8}
\fmf{phantom,tension=.5}{o1,v1}
\fmf{phantom,tension=.5}{o2,v2}
\fmf{phantom,tension=.5}{o3,v3}
\fmf{phantom,tension=.5}{o4,v4}
\fmf{phantom,tension=.5}{o5,v5}
\fmf{phantom,tension=.5}{o6,v6}
\fmf{phantom,tension=.5}{o7,v7}
\fmf{phantom,tension=.5}{o8,v8}
\fmfcyclen{plain,width=2,right=0.20}{v}{8} 
\fmffreeze
\fmf{phantom}{v5,c}
\fmf{plain,width=2}{c,v1}
\fmf{phantom}{o4,o45,o5,o55,o6}
\fmffreeze
\fmf{boson,width=.5}{o45,oo,o55}
\fmf{dashes,width=.5}{oo,v5}
\fmfv{label=$\alpha'$,label.angle=-90,label.dist=6}{v7}
\fmfv{label=$\alpha_e$,label.angle=-90,label.dist=6}{c}
\end{fmfgraph*}}}
$$
The momentum $\alpha$ is replaced by $\alpha'=\alpha\pm b/2$.
Explicitly we find
\begin{equation}\label{eq:N=2star-Verlinde}
[\mathcal  L_{1,0}\cdot \mathcal F](\alpha,\alpha_e)=
\sum_\pm H_\pm(\alpha,\alpha_e)
\mathcal F(\alpha\pm b/2,\alpha_e)\,,
\end{equation}
where
\begin{align}\label{eq:H-factors}
H_\pm(\alpha,\alpha_e)
=\frac{\Gamma(\pm 2 b (\alpha-Q/2))\Gamma(\pm 2b  (\alpha-Q/2)+b Q)}
{\Gamma(\pm 2 b (\alpha-Q/2) + b \alpha_e)\Gamma(\pm 2 b (\alpha-Q/2)- b \alpha_e+b Q)}\,.
\end{align}
See \cite{Alday:2009fs,Drukker:2009id} for many more examples of Verlinde operators in Liouville theory.

\subsection{Comparison with gauge theory}
\label{sec:comparison}

Let us now compare the CFT results with the gauge theory results in Section \ref{sec:localization}.
We will write $\vec\alpha=(\alpha_j)_{j=1}^{3g-3+n}$.

According the AGT correspondence \cite{Alday:2009aq}, extended to generic $b$ \cite{Hama:2012bg}, the $\mathbb S^4_b$ partition function (\ref{eq:S4b-pf}) of the $A_1$-theory 
is the Liouville correlation function on $C_{g,n}$:
\begin{equation}\label{eq:gauge-Liouville-PF}
Z_{\mathbb S^4_b}
=\int [d\alpha]C(\vec\alpha) \overline{\mathcal F(\vec\alpha)}
 \mathcal F(\vec \alpha)
\,.
\end{equation}
Here $C(\vec \alpha)$ is an appropriate product of the three-point functions and $\mathcal F(\vec \alpha)$ is the conformal block.
A conjecture put forward in \cite{Alday:2009fs,Drukker:2009id}, again generalized to arbitrary $b$, is that the expectation value of a loop operator $L(\vec p,\vec q)$ on $\mathbb S^4_b$ is given by the ``expectation value'' of the Verlinde operator $\mathcal L_\gamma$:
\begin{equation}\label{eq:gauge-Liouville-loop}
  \langle L(\vec p,\vec q)\rangle_{\mathbb S^4_b} =
\int [d\alpha]C(\vec \alpha) \overline{\mathcal F(\vec\alpha)}
\mathcal L_\gamma\cdot \mathcal F(\vec \alpha)
\,.
\end{equation}

In order to compare gauge theory and CFT, it is natural to adopt a different normalization of conformal blocks \cite{Gomis:2011pf,Ito:2011ea}:
\begin{equation}\label{eq:block-normalized}
  \mathcal B(\vec\alpha):= C(\vec\alpha)^{1/2}\mathcal F(\vec \alpha)\,.
\end{equation}
This normalization is also natural for the quantization of the Hitchin system and for the interpretation of the fusion move as an analog of the 6$j$-symbols \cite{Teschner:2012em,Vartanov:2013ima}.
The new block $\mathcal B$ is identified with $Z^\text{pole}$, which contains not only the instanton contributions, but also the one-loop contributions from the north pole.
We define the action of a Verlinde operator on $\mathcal B$ as
\begin{equation}
(\mathbb L_\gamma\cdot \mathcal B)(\vec\alpha):=
 C(\vec\alpha)^{1/2}
(c_0 \mathcal L_\gamma\cdot \mathcal F)(\vec \alpha)
\end{equation}
by absorbing $c_0$ into $\mathbb L_\gamma$.
In terms of $\mathcal B$ and $\mathbb L_\gamma$, (\ref{eq:gauge-Liouville-loop}) becomes
\begin{equation}\label{eq:gauge-Liouville-loop2}
  \langle L(\vec p,\vec q)\rangle_{\mathbb S^4} =
\int [d\alpha]\,\overline{\mathcal B(\vec\alpha)}\,
\mathbb L_\gamma\cdot \mathcal B(\vec \alpha)\,.
\end{equation}

The Verlinde operator corresponding to a Wilson operator is still multiplicative:
\begin{equation}
\mathbb L_W\cdot \mathcal B=(e^{2\pi  ba}+e^{-2\pi  ba})\mathcal B\,.
\end{equation}
For Verlinde operators that involve shifts in 
$\vec\alpha$, the action is modified.
For example, the Verlinde operator $\mathbb L_{1,0}$ on the one-punctured torus computed in (\ref{eq:N=2star-Verlinde}) and (\ref{eq:H-factors}) becomes
\begin{align}
\mathbb L_{1,0}
=
\sum_{s=\pm 1} e^{ i\frac{s}{4} b\partial_{ a}}
\left(
\prod_\pm
\frac{
{ \cosh ( 2\pi b a \pm \pi b m)
}}
{
 \sinh(\pm 2\pi b a +  \frac{\pi}{2}i b^2)
}
\right)^{1/2}
 e^{i\frac{s}{4} b\partial_{ a}}\,,
\label{eq:2star-L10}
\end{align}
where  $\alpha=Q/2+i a$, $\alpha_e=Q/2+i m$.

For the pure Wilson operator, the equality (\ref{eq:gauge-Liouville-loop2}) immediately follows from Pestun's computation of the Wilson loop vev \cite{Pestun:2007rz} and its generalization \cite{Hama:2012bg} reviewed in Section \ref{sec:localization}.
One can also confirm that the gauge theory result (\ref{eq:2star-thooft}) for the minimal 't Hooft loop in $\mathcal N=2^*$ theory agrees with the CFT result (\ref{eq:gauge-Liouville-loop2}) combined with (\ref{eq:2star-L10}) for $b=1$.
The expected expression (\ref{eq:S3b-thooft}) for the 't Hooft loop on $\mathbb S^4_b$ is consistent with (\ref{eq:gauge-Liouville-loop2}).
Many more tests of the correspondence were made in \cite{Drukker:2009id,Ito:2011ea}.

The exact calculation of the disorder operators such as 't Hooft loops, and its verification by independent methods, is one of the important advances that became possible by localization and the 2d-4d correspondence.

\subsection{Higher rank gauge groups and Toda theories}
\label{sec:Toda}

The higher rank ($N>2$) $A_{N-1}$-type theories of class $\mathcal S$ are quiver theories that involve $SU(n)$ gauge groups ($2\leq n\leq N$) as well as non-Lagrangian theories whose non-Abelian flavor symmetries are gauged \cite{Gaiotto:2009we,G13}.
In a higher rank extension of the AGT relation \cite{Alday:2009aq,Wyllard:2009hg}, these theories correspond to the $SU(N)$ Toda CFT on a Riemann surface $C$.
Liouville theory is a special case ($N=2$) of Toda theory.
The Toda theory possesses an extended chiral algebra, the $W_N$-algebra.

The complete dictionary between gauge theory line operators and geometric objects on $C$ has not been developed yet for $N>2$.
Still, for simple theories such as $SU(N)$ $\mathcal N=2^*$ theory and the $SU(N)$ theory with $N_\text{F}=2N$ flavors, one expects from brane constructions that the dictionary for minimal Wilson and 't Hooft operator is essentially the same as in the $N=2$ case.
The $W_N$-algebra possesses representations with various degeneracy conditions.
Using the so-called semi-degenerate fields $V_\mu$, one can construct Verlinde operators \cite{Drukker:2010jp,Passerini:2010pr,Gomis:2010kv}.

 For the Wilson loops in the fundamental and anti-fundamental representations, the Verlinde operator was calculated in \cite{Passerini:2010pr} by the monodromy of a degenerate block given in terms of a generalized hypergeometric function, of which (\ref{eq:block-hyper}) is a special case.
 The Verlinde operators for the minimal 't Hooft loops in $\mathcal N=2^*$ theory and the conformal SQCD were computed in \cite{Gomis:2010kv}, by determining the relevant fusion move matrices from the monodromy of generalized hypergeometric functions.
The agreement between the 4d and 2d results reviewed in Section \ref{sec:comparison} extends to the higher rank case.

The authors of \cite{Drukker:2010jp} expressed the Verlinde operator for the Wilson loop curve in terms of the fusion and braiding moves.
General identities that follow from the axioms of CFT imply that such a Verlinde operator inserts
$S_{\mu,\alpha}/S_{0,\alpha}$ into the Toda version of (\ref{eq:gauge-Liouville-PF}), where $\mu$ is the semi-degenerate representation, $\alpha$ is the generic representation propagating across the curve, and $S$ denotes the modular S-matrix.
(See also  \cite{Wu:2009tq}.)
This turns out to be the same as the insertion of a so-called topological defect along the curve.
The latter is a one-dimensional object in CFT, defined by the condition that the holomorphic and anti-holomorphic generators of the chiral algebra commute with it.
Topological defects were originally constructed in \cite{Petkova:2000ip} for rational CFT and in \cite{Sarkissian:2009aa} for Liouville theory.
Since the Verlinde operators and the topological defects transform in the same way under the modular groupoid, they must be identical.
The definition of a topological defect is local and does not require introducing conformal blocks.
It is also possible to compute 't Hooft loops using topological defects in Liouville theory \cite{Petkova:2009pe}.
Another feature of a higher-rank Toda theory is that Verlinde operators/topological defects can be defined on networks with trivalent vertices \cite{Drukker:2010jp,Bullimore:2013xsa}.
The identification of the corresponding line operators is an interesting open problem.

\section{Line operator algebras and the Hitchin moduli space}
\label{sec:algebra}

The previous two sections concerned the expectation values and correlation functions of line operators.
It turns out that the line operators also possess interesting algebraic structures.

\subsection{Operator product expansion from SUSY quantum mechanics}
\label{sec:algebra-SQM}

Let us consider a general $\mathcal N=2$ theory on $\mathbb R\times I\times\mathcal C$, where $I$ is an interval and $\mathcal C$ (not to be confused with $C$) is a  Riemann surface.
With a suitable twist along $\mathcal C$, the theory depends on the complex structure of $\mathcal C$ but not on its K\"ahler structure \cite{Johansen:1994aw,Kapustin:2006pk}.
It is also independent of the gauge coupling and can be analyzed at weak coupling.
 A Wilson-'t Hooft operator along $\mathbb R$, inserted at a point $(s,z)\in I\times\mathcal C$, is annihilated by a fermionic charge $\mathcal Q$ which is a scalar along $\mathcal C$.
 The correlator of two such operators $L_i$ ($i=1,2$) depends holomorphically on the complex coordinates $z_i$ on $\mathcal C$, and is (locally) independent of the positions $s_i$ on $I$.
We impose suitable boundary conditions at the two ends of $I$.

 At low energies, the theory reduces to an $\mathcal N=2$ quantum mechanics whose target space is the moduli space of solutions of the Bogomolny equations on $I\times\mathcal C$, possibly with Dirac monopole singularities of charge $B_i$ at $(s_i,z_i)$.
The data for $\mathcal N=2$ quantum mechanics also include a holomorphic vector bundle determined by the electric charges of $L_i$, as follows from the construction of dyonic operators in Section \ref{sec:line-def}.
In simple cases the moduli space can be described rather explicitly.
The BPS Hilbert space is the $L^2$ Dolbeault cohomology of the vector bundle.

Let us set $z_1=z_2$ and take $s_1\neq s_2$.
The moduli space in the limit $s_1\rightarrow s_2$ develops a singularity.
In simple examples, the singularity corresponds to shrinking exceptional divisors.
The $L^2$ cohomology then splits into several parts, one with elements localized to the smooth part of the moduli space, and the others with support in the vicinity of a divisor.
The latter correspond to smaller magnetic charges, and is a manifestation of the phenomenon``monopole bubbling'' \cite{Kapustin:2006pk}.

This method was applied in \cite{Kapustin:2006pk,Kapustin:2006hi,Kapustin:2007wm,Saulina:2011qr,Moraru:2012nu} to compute the operator product expansion (OPE) of line operators in the form
\begin{equation}
  L(B_1,w_1)\cdot L(B_2,w_2)=L(B_1+B_2,w_1+w_2)+\sum_j (-1)^{s_j} L(B_j,w_j)\,,
\end{equation}
where $(-1)^{s_j}$ are signs, and $(B_k,w_k)\in  \Lambda_{\rm m}\times\Lambda_{\rm w}$ in the notation of Section \ref{sec:line-def}.
The signs arise because we weight the BPS Hilbert space by the fermion number.
Various checks have been made by S-duality.

The results in \cite{Moraru:2012nu} should be compared with the two-dimensional methods developed in \cite{Xie:2013lca,Xie:2013vfa,Bullimore:2013xsa,Cirafici:2013bha,Saulina:2014dia} for higher-rank theories.
It appears that more work is needed to have a unified view on the algebra of line operators for higher-rank class $\mathcal S$ theories.

\subsection{Non-commutative algebra of line operators}
\label{sec:algebra-S1R3}

The set-up $\mathbb R\times I\times \mathcal C$ above can be identified with $\mathbb S^1\times_b\mathbb R^3$ ($=\mathbb S^1\times_b\mathbb R\times\mathbb R^2$)  without omega deformation ($b=0$) by the identification $(\mathbb R, I, \mathcal C)\rightarrow (\mathbb S^1, \mathbb R, \mathbb R^2)$.
Recall also that the latter geometry with general $b$ is the effective geometry in the neighborhood of the loop operator in $\mathbb S^4_b$ and $\mathbb S^1\times_b \mathbb S^3$.
Thus the algebra of line operators on $\mathbb S^1\times_b\mathbb R^3$ captures the counterparts in other geometries.

The OPE of two operators on $\mathbb S^1\times_b(\mathbb R\times\mathbb R^2)$ with $b\neq 0$ depends on the ordering along the $\mathbb R$.
Namely, $L_1\cdot L_2$ (for $s_1>s_2$) in general does not equal $L_2\cdot L_1$ (for $s_1<s_2$).
This is because the Poynting vectors in the two cases contribute to the trace (\ref{eq:S1R3-trace}) with opposite signs of angular momentum $J_3$ \cite{Gaiotto:2010be}.
The OPE takes the form
\begin{equation}\label{eqn:OPE-omega}
  L(B_1,w_1)\cdot L(B_2,w_2)=e^{s_{12}\pi i b^2} L(B_1+B_2,w_1+w_2)+\sum_j c_j(b) L(B_j,w_j)\,,
\end{equation}
where $s_{12}=\langle B_2,w_1\rangle-\langle B_1,w_2\rangle$ is a symplectic pairing and the coefficients $c_j$ ($j\neq 1,2$) depend on $b$.
In fact the localization analysis shows that $\langle L_1\cdot L_2\rangle_{\mathbb S^1\times_b\mathbb R^3}=\langle L_1 \rangle_{\mathbb S^1\times_b\mathbb R^3}*\langle L_2\rangle_{\mathbb S^1\times_b\mathbb R^3}$, where $*$ is the Moyal product:
\begin{equation}
(f*g)(\boldsymbol a, \boldsymbol b)\equiv 
\left.
e^{i\frac{b^2}{4\pi}
(
\partial_{\boldsymbol b}\cdot \partial_{\boldsymbol a'}
-
\partial_{\boldsymbol a}\cdot \partial_{\boldsymbol b'}
)
}f(\boldsymbol a,\boldsymbol b) g(\boldsymbol a',\boldsymbol b')
\right|_{\boldsymbol a'=\boldsymbol a,\boldsymbol b'=\boldsymbol b}\,.
\end{equation}
This product is associative but non-commutative, and is associated with the holomorphic symplectic structure $\Omega=d\boldsymbol a\wedge d\boldsymbol b$ with $\hbar=b^2/2\pi$ \cite{Ito:2011ea}.
Also, the relation between $\mathbb S^4_b$ and $\mathbb R^1\times_b \mathbb R^3$, together with the AGT correspondence, suggests that the corresponding Verlinde operator $\mathbb L$ acting on the normalized conformal blocks $\mathcal B$ in (\ref{eq:block-normalized}) is the Weyl ordering of $\langle L\rangle_{\mathbb S^1 \times_b\mathbb R^3}$ viewed as a function of $\boldsymbol a$ and $\boldsymbol b$ \cite{Ito:2011ea}.

\subsection{Quantization of the Hitchin moduli space}
\label{sec:Hitchin}

The various 4d geometries considered in Section \ref{sec:localization} admit a natural action of $U(1)\times U(1)$ isometries.
In the tubular neighborhood of a supersymmetric loop operator, or two such operators very close to each other, the local geometry can be effectively approximated by $\mathbb S^1\times_b\mathbb R^3$, where one $U(1)$ rotates the $\mathbb S^1$ and the other acts as rotations about the 3-axis.
Since the algebra of supersymmetric line operators is a UV property of the theory, it suffices to analyze the theory on $\mathbb S^1\times_b \mathbb R^3$.

For class $\mathcal S$ theories, line operators are intimately related to functions on the Hitchin moduli space.
One can see this most clearly as follows \cite{Gaiotto:2010be}.
Recall that a class $\mathcal S$ theory is specified by a simply laced Lie algebra $\mathfrak g$ and a Riemann surface $C$ \cite{Gaiotto:2009we,Gaiotto:2009hg,G13}.
For simplicity we assume that there is no puncture.
For $b=0$ such a theory on $\mathbb S^1\times_b\mathbb R^3$ is simply the 6d $\mathcal N=(0,2)$ theory on $\mathbb S^1\times\mathbb R^3\times C$, topologically twisted along infinitesimally small $C$.
If instead the size of $C$ is much bigger than the radius of the $\mathbb S^1$, a better description is the 5d maximally supersymmetric Yang-Mills theory on $\mathbb R^3\times C$.
The condition for preserving supersymmetry is precisely the Hitchin equations on $C$ \cite{Bershadsky:1995vm,Gaiotto:2010be}:
\begin{equation}
F_{z\bar z}=[\varphi_z,\bar\varphi_{\bar z}]\,,    \quad
 D_{\bar z}\varphi_z=0\,,
\quad 
 D_z \bar \varphi_{\bar z}=0\,,
\end{equation}
where $\varphi_z$ and its complex conjugate $\bar\varphi_{\bar z}$ arise from two real scalars via twisting.
The space of solutions modulo gauge transformations taking values in the simply connected group $\tilde G$ with Lie algebra $\mathfrak g$ is the Hitchin moduli space $\mathcal M_H(\tilde G,C)$.
The Coulomb moduli space of the 5d theory on $\mathbb R^3\times C$ is believed to be the quotient $\mathcal M_H(\tilde G ,C)/\Delta$ by a discrete group $\Delta$.
Here $\Delta$ is a subgroup of the group of flat line bundles whose structure group is the center $\mathcal C$ of $\tilde G$, and can be identified with the maximal coisotropic subgroup of $H^1(C,\mathcal C)$ denoted by the same symbol in Section \ref{sec:spectrum} \cite{Gaiotto:2010be,Tachikawa:2013hya}.
The line operators in the 4d theory arise from surface operators in the 6d theory wrapping a curve (or a trivalent network) on $C$.
The surface operators descend to Wilson loops for a complex gauge field obtained from $(A_z,\varphi_z)$  in the 5d theory. 
We note that the twist along $C$ eliminates dependence of BPS observables on the scale of the metric on $C$, and that the 5d theory is IR free.
It is then natural to expect \cite{Gaiotto:2010be} that the correlation functions of BPS line operators on $\mathbb S^1\times_{b=0}\mathbb R^3$ are given by the classical holonomies on $C$.
This expectation was shown to be consistent with wall-crossing \cite{Gaiotto:2010be} in the 4d IR theories, and was also directly demonstrated \cite{Ito:2011ea} for a few examples by noting that the parameters $(\boldsymbol a,\boldsymbol b)$ in Section \ref{sec:S1R3} are the complexification of the Fenchel-Nielsen coordinates on the Hitchin moduli space.

In order to see that the omega deformation $\mathbb S^1\times_b\mathbb R^3$ induces non-commutativity, one approach is to reduce the theory by the action of $U(1)\times U(1)$ to two dimensions \cite{Nekrasov:2010ka}.
This can be done for a topologically twisted theory, and in the limit that the orbits of the action become small,  the reduced 2d theory is the $\mathcal N=(4,4)$ sigma model with target space $\mathcal M_H(\tilde G ,C)/\Delta$.
The 4d geometry reduces to a half plane, and line operators get inserted along the boundary.
The presence of a B-field accounts for non-commutativity \cite{Kapustin:2005vs}.

If we reduce $\mathbb S^4_b$ by the action of $U(1)\times U(1)$ above, the neighborhood of the equator $\mathbb S^3_b$ of $\mathbb S^4_b$ reduces to a two-dimensional strip, as considered in \cite{Nekrasov:2010ka}.
By topologically twisting the 4d theory, one obtains a two-dimensional sigma model.
The line operators along the circle $\mathbb S^1_{(b)}$ define the boundary chiral ring $\mathcal A_b$ on the left boundary, while those along $\mathbb S^1_{(1/b)}$ define another ring $\mathcal A_{1/b}$ on the right boundary.
The $A_1$-theory on $\mathbb S^4_b$ realizes the quantization of the Hitchin moduli space with a Hilbert space; the two rings act on the Hilbert space of conformal blocks.
If we included all the operators labeled by $\Lambda_{\rm m}\times \Lambda_{\rm w}/(\text{Weyl group})$, $\mathcal A_b$ and $\mathcal A_{1/b}$ would not commute because the two circles $\mathbb S^1_{(b)}$ and $\mathbb S^1_{(1/b)}$ are linked inside the $\mathbb S^3_b$ in the constant time slice $\{X_0=0\}$  \cite{'tHooft:1977hy}, as explained in Section \ref{sec:spectrum}:
\begin{equation}
L_{\gamma_1}^{(b)}\cdot L_{\gamma_2}^{(1/b)}=(-1)^{\langle \gamma_1,\gamma_2\rangle}  L_{\gamma_2}^{(1/b)}\cdot L_{\gamma_1}^{(b)}\,,
\end{equation}
where we denote by $\gamma_j$ the corresponding charges $(\vec p_j,\vec q_j)$, and $\langle \gamma_1,\gamma_2\rangle$ is a symplectic product.
Indeed as explained in Section \ref{sec:spectrum}, we must restrict to a maximal mutually local subset of  $\Lambda_{\rm m}\times \Lambda_{\rm w}/(\text{Weyl group})$ such that $\langle \gamma_1,\gamma_2\rangle$ is even for any pair of line operators.
Then $\mathcal A_b$ and $\mathcal A_{1/b}$ commute with each other, as they should because they are chiral rings on two separate boundary components.
As explained above such a restriction modifies the target space from $\mathcal M_H(C,SU(2))$ to its quotient by a finite group $\Delta$ \cite{Gaiotto:2010be}.

These are manifestations of the relation between the $A_{N-1}$-gauge theories and quantization of $\mathcal M_H(C,SU(N))$ associated with the curve $C$.
See \cite{Ne} for discussions and references.
The connection between the gauge theory and the Hitchin system can also be used to study line operators from the IR point of view, where a different class of Darboux coordinates naturally appears \cite{Gaiotto:2010be,Ne}.
For some theories the non-commutative algebra of line operators can be computed
using IR quiver quantum mechanics \cite{Cordova:2013bza}.
An important open problem is the comparison of the algebraic relations obtained in different approaches.

\section*{Acknowledgements}

I am grateful to my collaborators for enlightening discussions on this subject.
I also thank N.~Drukker and the referee in this project for useful comments on drafts.
This research is supported in part by Grant-in-Aid for Young Scientists (B) No. 23740168 and by Grant-in-Aid for Scientific Research (B) No. 25287049.

\appendix

\section{Summary of relevant facts}
\label{app:summary}

\subsection{Class $\mathcal S$ theories of type $A_1$}
\label{app:class-S}

The low-energy theory in the world-volume of two M5-branes is a six-dimensional  ${\mathcal N}=(0,2)$ supersymmetric theory with no known Lagrangian description.
An $A_1$-theory of class $\mathcal S$ is believed to arise by compactifying the six-dimensional theory on the Riemann surface $C_{g,n}$ of genus $g$ with $n$ punctures, with each puncture carrying a codimension-two defect of the $(0,2)$ theory \cite{Gaiotto:2009we,Gaiotto:2009hg,G13}.
The $A_1$-theories of class $\mathcal S$ provide basic examples of 2d-4d correspondence.

A weakly coupled description of such a theory may be encoded in a choice of decomposition of $C_{g,n}$ into $3g-3+n$ pairs of pants, and a trivalent graph $\Gamma$ drawn on $C_{g,n}$.
Each pair of pants contains one vertex, and three edges come out through distinct boundary components (pants legs).
An example is shown in Figure \ref{fig:torus}(a).
We allow a pants leg to degenerate to a puncture.
The graph $\Gamma$ has $3g-3+n$ internal edges and $n$ external edges ending on the punctures.
The field content in this description of the ${\cal N}=2$ theory can be read off from $\Gamma$ by associating to each internal edge an $SU(2)$ gauge group and to each vertex eight half-hypermultiplets in the trifundamental representation of the $SU(2)^3$ group associated to the three attached edges. 
When the edge is external the $SU(2)$ symmetry corresponds to a flavor symmetry.
A change of pants decomposition and $\Gamma$ corresponds to a S-duality transformation.

\subsection{Liouville theory}
\label{app:Liouville}

Liouville field theory is formally defined by the path integral over a single real field $\phi$ weighted by $e^{-S}$, where
\begin{equation}
  S=\frac{1}{4\pi} \int_C  \left(\partial^\mu\phi \partial_\mu\phi +4\pi \mu e^{2b\phi}+QR \phi\right)\,.
\end{equation}
Here 
$R$ is the scalar curvature, and $Q=b+1/b$ parametrizes the central charge $c=1+6Q^2$.
The ``cosmological constant'' $\mu$ can be absorbed into a shift of $\phi$, and affects the theory in a very mild way.
Liouville theory is a non-rational CFT, meaning that it contains infinitely many representations of the Virasoro algebra.
The spectrum of representations is continuous, and the conformal dimension $\Delta$ is parametrized by the Liouville momentum $\alpha \in Q/2+i\mathbb R_{\geq 0}$ as $\Delta=\alpha(Q-\alpha)$.
We denote the corresponding primary field by $V_\alpha$.

Fusion move coefficients $F_{s_1 s_2}=F_{s_1 s_2}
\big[\begin{smallmatrix} \alpha_3 & -b/2\\ \alpha_4 & \alpha_1\end{smallmatrix}\big]$ in (\ref{eq:fusion}) are explicitly known:
\begin{equation}\label{eq:fexp}\begin{aligned}
F_{++}&=\frac{\Gamma(b(2\alpha_1-b))\Gamma(b(b-2\alpha_3)+1)}{\Gamma(b(\alpha_1-\alpha_3-\alpha_4+b/2)+1)\Gamma(b(\alpha_1-\alpha_3+\alpha_4-b/2))}\,, \\
&\quad\,\text{etc.}
\end{aligned}
\end{equation}

\end{fmffile}

\paragraph{\large References to articles in this volume}
\renewcommand{\refname}{

\paragraph{Other references}
\renewcommand{\refname}{\vskip-36pt}


\bibliography{refs}

\end{document}